\documentclass[structabstract]{aa}

\usepackage[tight]{subfigure}
\usepackage{graphicx}
\usepackage{txfonts}
\usepackage{color}

\usepackage{natbib}
\bibpunct{(}{)}{;}{a}{}{,}

\begin{document}
   \title{The Rotation of Surviving Companion Stars after Type Ia Supernova Explosions in the WD+MS Scenario}

   \author{Zheng-Wei. Liu
          \inst{1,2,3,4},
           R. Pakmor
          \inst{5,4},
           F. K. R\"opke
          \inst{6,4},
           P. Edelmann
          \inst{4},
           W. Hillebrandt 
          \inst{4},\\
           W. E. Kerzendorf
          \inst{7,8}, 
           B. Wang
          \inst{1,2}
          \and
           Z. W. Han
          \inst{1,2}
          }

   \institute{National Astronomical Observatories/Yunnan Observatory, Chinese Academy of Sciences, Kunming 650011, P.R. China\
         \and
              Key Laboratory for the Structure and Evolution of Celestial Objects, Chinese Academy of Sciences, Kunming 650011, P.R. China\
         \and
              University of Chinese Academy of Sciences, Beijing 100049, P.R. China\
         \and
              Max-Planck-Institute f\"ur Astrophysik, Karl-Schwarzschild-Str. 1, 85741 Garching, Germany\\
             \email{zwliu@ynao.ac.cn}    
         \and
              Heidelberger Institut f\"ur Theoretische Studien, Schloss-Wolfsbrunnenweg 35, 69118 Heidelberg, Germany\
         \and
              Institut f\"ur Theoretische Physik und Astrophysik, Universit\"at W\"urzburg, Am Hubland, 97074 W\"urzburg, Germany\
         \and
             Research School of Astronomy and Astrophysics, Mount Stromlo Observatory, Cotter Road, Weston Creek, ACT 2611, Australia\
         \and
            Department of Astronomy and Astrophysics, University of Toronto, 50 Saint George Street, Toronto, ON M5S 3H4, Canada\\
            }

  \abstract
   {In the single degenerate (SD) scenario of type Ia supernovae (SNe Ia)
 the non-degenerate companion star survives the supernova (SN)
 explosion and thus should be visible near the center of the SN
 remnant and may show some unusual features. Therefore, a promising
 approach to test progenitor models of SNe Ia is to search for the
 companion star in historical SN remnants.}
   {Here we present the
 results of three-dimensional (3D) hydrodynamics simulations of the 
 interaction between the SN Ia blast wave and a MS companion taking into 
 consideration its orbital motion and spin. The primary goal of this work
 is to investigate the rotation of surviving companion stars after SN Ia
 explosions in the WD+MS scenario.}
   {We use Eggleton's stellar evolution code including the optically
    thick accretion wind model to obtain realistic 
    models of companion stars. The impact of the supernova blast
     wave on these companion stars is followed in 3D
     hydrodynamic simulations employing the smoothed particle
     hydrodynamics (SPH) code GADGET3.}
   {We find that 
 the rotation of the companion star does not significantly affect the 
 amount of stripped mass and the kick velocity caused by  the SN
 impact. However, in our simulations, the rotational velocity 
 of the companion star is significantly reduced to about $14\%$ to $32\%$ of 
 its pre-explosion value due to the expansion of the companion and 
 the fact that $55\%-89\%$ of the initial angular momentum is carried 
 away by the stripped matter. }
   {Compared with the observed 
 rotational velocity of the presumed  companion star of Tycho's
 supernova, Tycho G, of $\sim 6\,\rm{km\,s^{-1}}$ the final
 rotational velocity we obtain in our simulations is still
 higher by at least a factor of two. Whether or no this difference is
 significant, and may cast doubts on the suggestion that Tycho G is the
 companion of SN 1572, has to be investigated in future studies. 
 Based on binary population synthesis
 results we present, for the first time, the expected distribution 
 of rotational velocities of companion stars after the  SN explosion
 which may provide useful information for the identification of the 
 surviving companion in observational searches in other historical 
 SN remnants. }


   \authorrunning{Z. W. Liu et al.}
   
   \titlerunning{Post-explosion rotation of surviving companions}

   \maketitle
%

\section{Introduction}
  Type Ia SNe (SNe Ia) are used as cosmic distance indicators since
  their luminosity can be calibrated based on the empirical relation 
  between light curve shape and peak luminosity \citep{Phil93, Phil99}. 
  This has provided the first evidence for the accelerating expansion
  of the present universe \citep{Ries98, Perl99, Leib08}. 
  SNe Ia are also believed to be important contributors to the
  cosmic nucleosynthesis and they are sources of kinetic energy in 
  evolution process of galaxies. Although a detailed understanding of 
  the nature of their progenitors and the physics of explosion is 
  still lacking (see \citealt{Hill00} for a review), there 
  is consensus that SNe Ia arise from thermonuclear explosions of 
  carbon/oxygen white dwarfs (CO WDs) in binary systems \citep{Hoyl60, Nomo97}.

  In principle, there are various possibilities for the evolution
  towards an explosion (see \citealt{Wang12} for a review). One option is a merger
  of two CO WDs with a combined mass in excess of the Chandrasekhar
  mass, which may explode as a SN Ia, the `double degenerate' (DD)
  scenario \citep{Iben84, Webb84}.  The DD model can explain in a
  natural way the lack of hydrogen in SNe Ia. However, only a few DD
  systems have been found whose orbital period is short enough that
  they will merge within a Hubble time, but in none of them the
  combined mass exceeds the Chandrasekhar mass limit. On the other
  hand, in recent numerical simulations of \citet{Pakm10, Pakm11} it
  was found that the violent merger of a pair of white dwarfs with
  equal masses of $\sim 0.9\,\mathrm{M}_{\odot}$ can directly trigger a
  thermonuclear explosion that resembles sub-luminous 1991bg-like SNe
  Ia. Moreover, it was shown that the violent merger of two CO WDs
  with masses of $0.9\,\mathrm{M}_{\odot}$ and $1.1\,\mathrm{M}_{\odot}$
  produces lightcurves and spectra which are in good agreement with
  those of normal SNe Ia \citep{Pakm12}, supporting the DD scenario.

  On the other hand, a rather massive WD may accrete hydrogen-rich 
  material from a non-degenerate binary companion until it
  approaches the Chandrasekhar mass, the `single degenerate' (SD)
  scenario. In this case, the binary companion could be a
  main-sequence (MS) star (WD+MS channel), a slightly evolved subgiant
  star or a red-giant (RG) star \citep{Han04, Wang10a}.  The lack of
  hydrogen in observed SN Ia spectra can be seen as troublesome for
  the SD scenario since the companions are hydrogen-rich stars. With
  more realistic MS companion star models than those used in previous
  work, \citet{Liu12} performed three-dimensional (3D) smooth particle
  hydrodynamics (SPH) simulations of the interaction between SN Ia
  ejecta and the MS companion star. They found that in all cases they
  considered more than $0.1 \mathrm{M}_{\odot}$ was stripped from the
  hydrogen-rich companion and was mixed into the SN ejecta, which is
  in disagreement with most recent observational constraints on the
  presence of hydrogen in SN Ia from nebular spectra $\sim
  0.01\,\mathrm{M}_{\odot}$ \citep{Leon07, Shap12}.  Moreover, no other
  similar hydrodynamics simulations showed an amount of stripped mass
  below this strong observational limit (see \citealt{Mari00, Pakm08,
    Pan10, Rick10, Pan12b}).  On the other hand, there is evidence
  from observations that the progenitors of at least some SNe Ia come
  from the SD channel. For instance, evidence for the presence of
  circumstellar matter and features indicative for an interaction
  between the SN and circumstellar matter were found recently
  \citep{Pata07, Ster11, Fole12, Dild12}.

  The merger of two WDs leaves no remnant after a SN Ia explosion. In
  contrast, in the SD scenario the companion star survives the
  explosion and, in principle, can be identified due to its peculiar
  spatial velocity, its rotation, effective temperature, luminosity or
  composition \citep{Han08, Wang10b}. Therefore, it is a promising
  approach to test the progenitors of SNe Ia by directly searching for
  the surviving companion star in galactic SN remnants (SNRs). There
  are several ways by which the SN blast wave modifies the properties
  of the companion. First, the SN strips off matter from the surface
  of the companion and injects thermal energy into it during the
  interaction. This causes the surviving companion to expand and
  lowers its surface gravity. Secondly, after the impact the companion
  star's surface will be enriched with heavy elements (e.g., Ni, Fe or
  Ca) from the inner part of the SN ejecta which should show up in 
  its spectrum (see \citealt{Gonz09, Pan12a, Pan12b}). Finally,
  after the explosion the companion star retains its pre-explosion
  orbital velocity, giving it a peculiar velocity compared to other
  stars in the vicinity.

  Tycho Brahe's SN 1572 is a SN Ia that exploded in the Milky
  Way. \citet{Ruiz04} have analyzed the stars within a circle of 0.65
  arcmin radius of the center of the SNR up to an apparent visual
  magnitude V = 22. They found a star, Tycho G, similar to the Sun in
  surface temperature and luminosity but with a lower surface gravity
  than a MS star. It has a significant peculiar velocity in radial and
  proper motion, and moves at more than three times the mean velocity
  of the other stars in the field. Therefore, they suggested that Tycho G
  star could be the surviving companion star of SN 1572. However,
  since then it has been noted that Tycho G does not show any spectral
  peculiarities \citep{Ihar07} and that it is apparently not out of 
  thermal equilibrium \citep{Howe11}. \citet{Fuhr05} also claimed that
  it might be a thick-disk star coincidentally passing in the vicinity
  of the remnant of SN 1572. Recently, \citet{Gonz09} found that Tycho
  G has an overabundance of Ni relative to normal metal-rich stars,
  and they suggest that Tycho G could have captured the
  low-velocity tail of the SN 1572 ejecta, which upholds Tycho G as a
  surviving companion star again. However, the measured $\rm{[Ni/Fe]}$
  ratio from a more recent study of \citet{Kerz12} seems to be not so
  unusual with respect to field stars with the same metallicity.

  In the SD case, because of the strong tidal coupling of a Roche-lobe
  filling donor, the donor star rotation is expected to be tidally
  locked to its orbital motion. This forces the binary donor star to
  have a spin corresponding to the orbital frequency of the binary system
  \citep{Han04, Kerz09}. After the SN explosion, the companion is
  released from its orbit and continues to rotate. Therefore, fast
  post-explosion rotation might be a signature of the donor star. The
  survivor could be in rapid rotation which would be measurable easily
  (see \citealt{Kerz09, Kerz12}). With the HIRES instrument on the
  Keck-I telescope and with Subaru high-resolution spectroscopy of
  star G in the Tycho SNR \citet{Kerz09, Kerz12} measured the
  rotational velocity of Tycho G and obtained only $\sim 6 \pm
  1.5\,\rm{km\ s^{-1}}$. Thus, they concluded that Tycho G is
  unlikely to be the surviving companion star of SN 1572 because it
  does not have the expected high rotational velocity \citep{Kerz09,
    Howe11}. However, the impact of the SN ejecta strips off material
  from the surface of the companion star, which can significantly
  reduce its angular momentum, thereby lowering the rotational
  velocity. Moreover, the SN impact and heating bloat the companion
  star, which will also slow down its spin.  Recently, \citet{Pan12a}
  computed the post-impact evolution of the remnant star produced in
  their multi-dimensional adaptive mesh refinement simulations of
  WD+MS-like models with the FLASH code \citep{Pan12b}. In line with
  the arguments given above, they conclude that Tycho G cannot be
  eliminated as being a promising progenitor candidate, based on its
  low rotational surface velocity only \citep{Pan12a}.

  In this paper, we present the results of 3D SPH simulations of the
  impact of SN Ia ejecta on a MS companion star including the orbital
  motion and spin of the companion.  The purpose of the work is to
  investigate how the SN Ia impact changes the rotation rate of the
  companion which will be useful for identifying the surviving
  companion star in SNRs. The paper is organized as follows. In
  Section~\ref{sec:code and model}, we describe the codes used and
  show an example of the initial companion star models used in our
  simulations. In Section~\ref{sec:effect} the effect of rotation on
  the interaction between SN Ia ejecta and its MS companion are
  discussed. Post-impact rotation features of a MS star are shown in
  detail in Section~\ref{sec:rotation}. Furthermore, the results of
  our simulations are compared with the observed rotation of star
  Tycho G in Section~\ref{sec:comparison}. Next in
  Section~\ref{sec:reestablished}, we discuss the distribution of the
  rotational velocities of the remnant stars after thermal equilibrium
  is reestablished. Finally, in Section~\ref{sec:conclusions}, we
  summarize our results.

\section{Numerical method and model}
\label{sec:code and model}

\subsection{Numerical codes and initial setup}
\label{sec:code}

   In order to obtain consistent MS companion star models at the onset
   of the SN explosion, similar to \citet{Liu12}, we used Eggleton's
   one-dimensional (1D) stellar evolution code \citep{Eggl71, Eggl72,
     Eggl73} to follow the binary evolution of a SD
   progenitor system in detail. Roche-lobe overflow (RLOF) is treated in the
   code as described by \citet{Han04}. The opacity tables used in our
   calculations are compiled by \citet{Chen07} from \citet{Igle96} and
   \citet{Alex94}. We use a typical Population I composition with
   hydrogen abundance X = 0.70, helium abundance Y = 0.28, and
   metallicity Z = 0.02. We set $\alpha = l/H_{\rm{P}}$, the ratio of
   mixing length to the local pressure scale height, to 2 and
   set the convective overshooting parameter $\delta_{\rm{ov}}$ to
   0.12 \citep{Pols97, Schr97}, which roughly corresponds to an
   overshooting length of $\sim 0.25$ pressure scale height.

   Furthermore, the latest version of the SPH code GADGET-3
   \citep{Spri01, Spri05} is employed to simulate the impact of SN Ia
   explosions on their MS companions by including the orbital and spin
   velocities of companion stars. The GADGET code was originally used
   for cosmological simulations, but it has been modified to make it
   applicable to stellar astrophysics problems (see \citealt{Pakm12a})
   and it has been used successfully to capture the main dynamical
   effects of the SN impact on its companion star (see
   \citealt{Pakm08, Liu12}). In our simulations, we aim at determining
   the rotational velocity of the companion star after the
   impact. Therefore the fact that no matter leaves the computational
   domain and that momentum, energy and angular momentum are strictly
   conserved are a crucial advantage of SPH over grid-based methods
   for the problem under investigation.

   In this work, the basic setup for the GADGET code is almost the
   same as in \citet{Liu12}. The smoothing length is chosen such that
   a sphere of its radius encloses $60$ neighboring particles. All
   particles have the same mass. The gravitational softening length
   is equal to the smoothing length. To reduce numerical noise introduced
   by mapping the initial model, the MS companion star is relaxed for 1.0
   $\times$ 10$^4\,\rm{s}$ (several dynamical timescales) before we
   start the actual impact simulation. The W7 SN Ia model
   \citep{Nomo84} is used for the SN explosion, and the orbital
   separation is taken from the 1D consistent binary-evolution
   calculations.

\begin{figure}
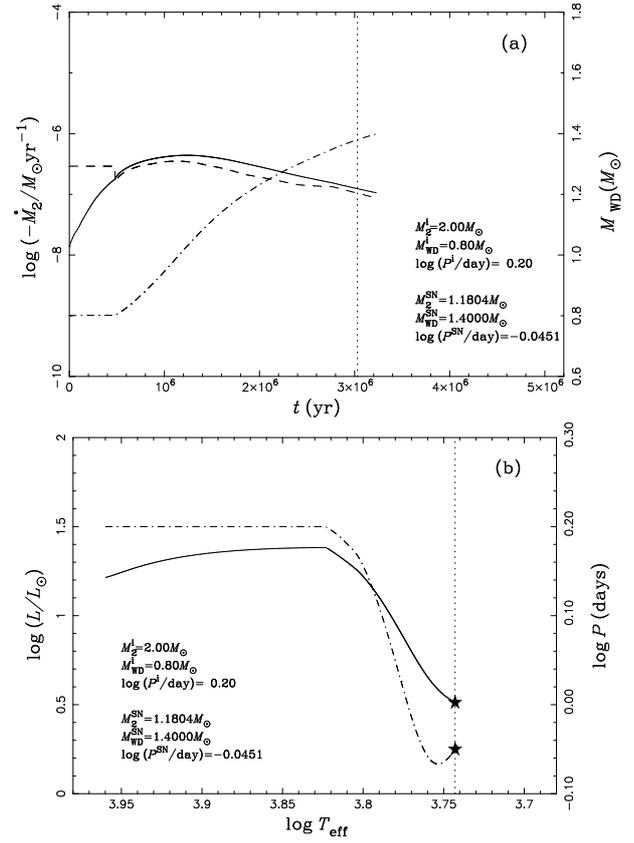

\centering
\includegraphics[width=0.3\textwidth, angle=270]{f1a.ps}
\includegraphics[width=0.3\textwidth, angle=270]{f1b.ps}
\caption{Binary-evolution calculation of the MS\_110 progenitor model. In panel(a), 
         the solid, dashed and dash-dotted curves show the mass-transfer rate,
         $\dot{M}_{\rm{2}}$, the mass-growth rate of the CO WD, $\dot{M}_{\rm{CO}}$, 
         the mass of the CO WD, $M_{\rm{WD}}$, respectively. In panel (b), the evolutionary
         track of the donor star is shown as a solid curve and the evolution of 
         the orbital period is shown as a dashed-dotted curve. The dotted vertical lines in both 
         panels and asterisks in panel (b) indicate the position where
         the WD explodes as a SN Ia.}
\label{Fig:hrd}
\end{figure}

\begin{figure*}
  \begin{center}
  \includegraphics[width=0.85\textwidth, angle=270]{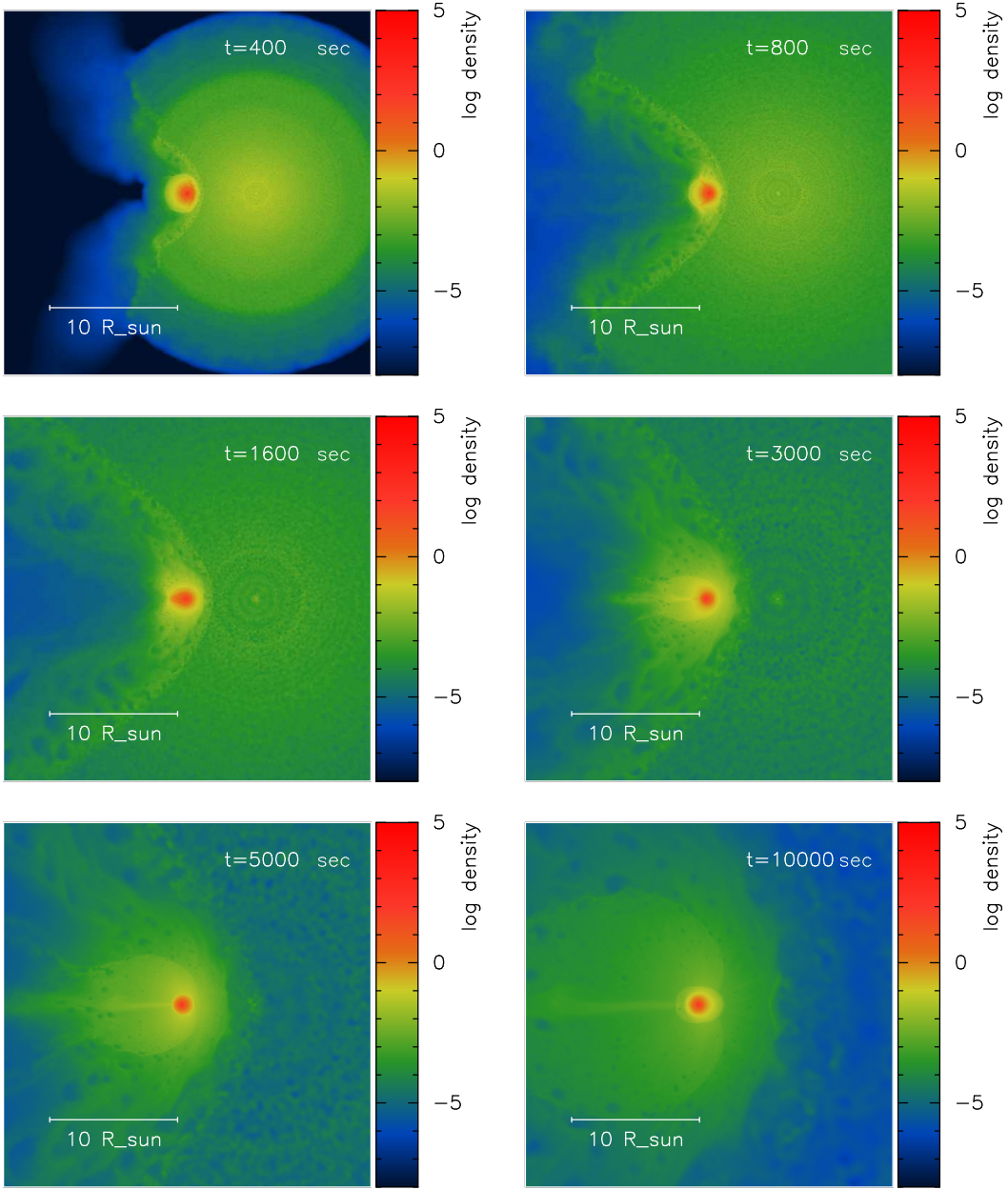}
              \caption{Density distribution of the SN ejecta and the companion star 
               of a non-rotating version of model MS\_110. The color
               scale shows the logarithm of the density. The plots are
               made using the freely available $SPLASH$ tool \citep{Price07}.}
    \label{Fig:rho}
  \end{center}
\end{figure*}

\begin{figure*}
  \begin{center}
  \includegraphics[width=0.85\textwidth, angle=270]{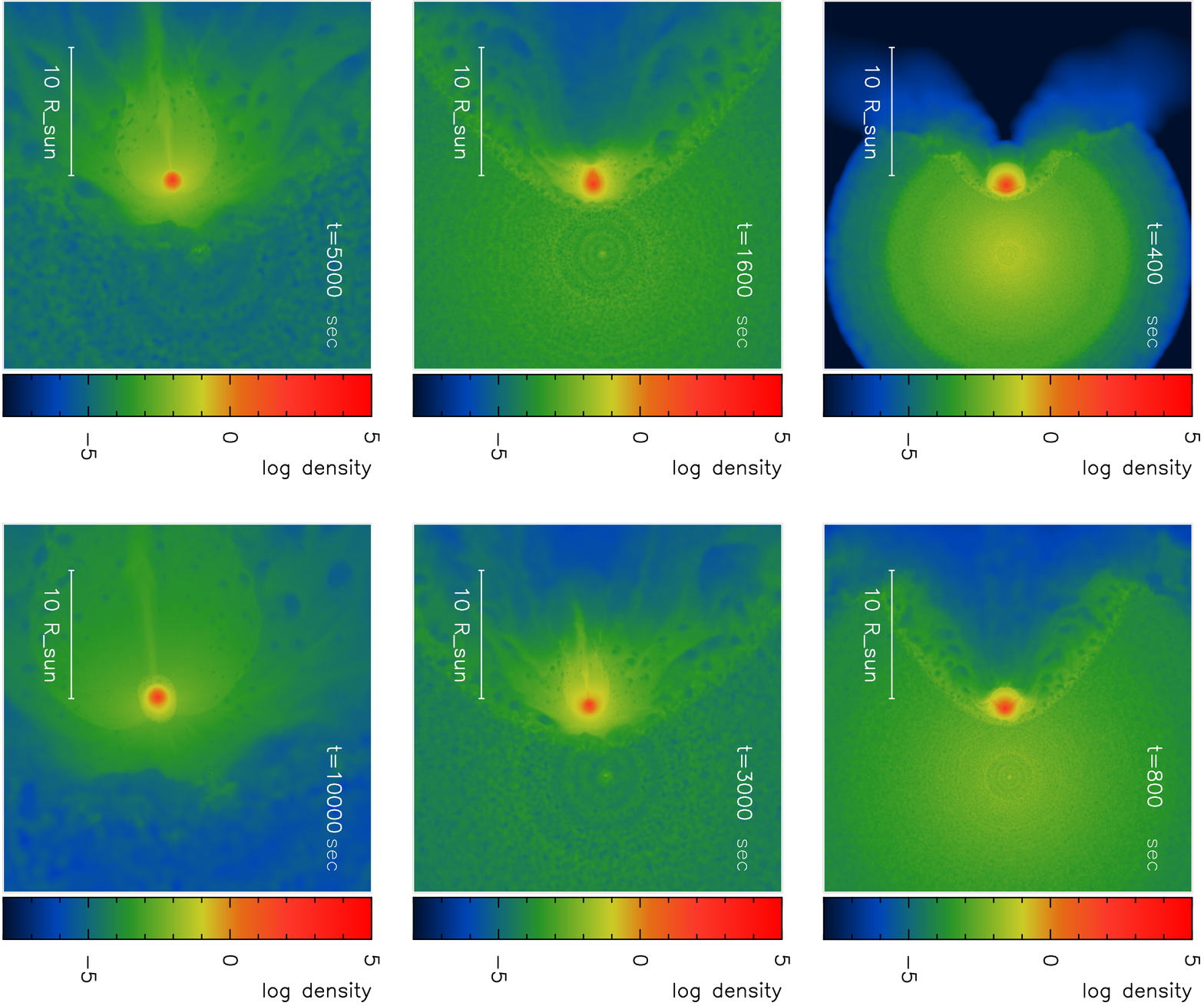}
    \caption{Same as Fig.~\ref{Fig:rho}, but including the orbital
             motion and the spin of the companion star. 
             The color scale shows the logarithm of the density. 
             Some differences become visible at late times.
             The plots are made using the freely available 
             $SPLASH$ tool \citep{Price07}.}
    \label{Fig:rho_rot}
  \end{center}
\end{figure*}

   Additionally, in this work, we assume that the companion star
   co-rotates with its orbit due to strong tidal interaction. Thus
   the spin period of the companion star is locked with its orbital
   period. Furthermore, the orbital and spin
   velocities of the companion stars are included in our impact
   simulations in order to study the post-impact rotation rate of
   surviving star. 

   The system of coordinates is chosen as follows. We set the $x-y$
   plane to be the orbital plane of the binary system and assume a
   circular orbit.  The $z-$axis is chosen as the spin axis of the
   companion, and the positive $z-$axis is the direction of the angular
   momentum. Finally, we assume that the companion star rotates as a
   rigid body at the moment of the SN explosion.

\subsection{Progenitor model}
\label{sec:model}

   As in \citet{Liu12} (see also \citealt{Han04, Wang10b}), we start
   our 1D consistent binary calculation when the WD+MS system has been
   formed, the mass transfer then occurs through RLOF when the
   companion star fills its Roche lobe. Instead of solving the stellar
   equations for the WD star when we trace the detailed evolution of
   the companion star, the optically thick wind model of
   \citet{Hach99} is adopted.

   Figure~\ref{Fig:hrd} shows the evolution of a binary system with an
   initial donor-star mass of $M^{\rm{i}}_{\rm{2}} =
   2.0\,\mathrm{M}_{\odot}$, an initial mass of the CO WD of
   $M^{\rm{i}}_{\rm{WD}} = 0.8\,\mathrm{M}_{\odot}$ and an initial orbital
   period of $\rm{log}(P^{\rm{i}}/\rm{day}) = 0.2$. In such a system,
   the companion star has an orbital velocity of
   $138\,\rm{km\ s^{-1}}$ at the moment of the SN Ia
   explosion. Assuming that the rotation of companion is locked
   with the orbital motion due to the strong tidal interaction, the
   rotational velocity, $v_{\rm{rot}}$, and its orbital velocity,
   $v_{\rm{orb,2}}$, obey a simple relation as follows (see also
   \citealt{Kerz09}):
\begin{equation}
   v_{\rm{rot}}=\frac{M_{\rm{1}}+M_{\rm{2}}}{M_{\rm{1}}}f(q)v_{\rm{orb,2}},
\end{equation}
 where
$q = M_{\rm{2}}/M_{\rm{1}}$ is the mass ratio of the binary system at 
the moment of the explosion, and $f(q)$ is the ratio of the Roche-lobe 
radius of companion star to the orbital separation  \citep{Eggl73}. 
Therefore, we calculate the rotational velocity of the companion star to be  
$v_{\rm{rot}}^{\rm{SN}} \sim 110\,\rm{km\ s^{-1}}$ at the time of the 
explosion.

In this model, because hydrogen burning on the WD is unstable before
the SN explosion, the system may be observed as a U Sco-type recurrent
nova \citep{Hach08, Meng10}. With our consistent binary-evolution
calculations, we obtain a CO WD+MS progenitor that is similar to the U
Sco binary system with a mass of the secondary of $1.18\, M_{\sun}$
and an orbital period $\sim 1\,\rm{day}$ (see Table~\ref{table:1}). This
model is named ``MS\_110'', the ``MS'' indicating a CO WD+MS binary
system, the ``110'' means the rotational velocity of the companion
star of $v_{\rm{rot}}^{\rm{SN}} \sim 110\,\rm{km\ s^{-1}}$ at the
moment of the explosion. U Scorpii is one of the best-observed
recurrent novae, and it has been suggested as a progenitor of
a SN Ia because its white dwarf mass is close to the Chandrasekhar
mass \citep{Hach00, Thor01, Pods03}. In order to investigate the
physical parameters of the recurrent nova, \citet{Hach00} have
successfully modeled the theoretical light curve for the outburst of
the U Scorpii system with a model consisting of a WD mass of $M_{\rm{WD}} =
1.37 \mathrm{M}_{\odot}$, a secondary mass of $M_{\rm{2}} = 0.8-2.0
\mathrm{M}_{\odot}$ and an orbital inclination of $\sim 80^{\circ}$.

\section{Effects of rotation}
\label{sec:effect}

Based on the system similar to U Sco (MS\_110) that was discussed in
Section~\ref{sec:model}, we performed a 3D SPH impact simulation which
included the spin velocity of $v_{\rm{rot}}^{\rm{SN}}=110\,
\rm{km\,s^{-1}}$, and the orbital velocity of
$v_{\rm{orb,2}}^{\rm{SN}} = 138\, \rm{km\,s^{-1}}$ to study the effect
of asymmetry due to the orbital motion and spin of the companion
star. We used a total of $2 \times 10^{7}$ particles to represent the
companion star only (which corresponds to a total number of particles
in the simulation of $\sim 4 \times 10^{7}$).  All particles had the
same mass of $5.9 \times 10^{-8}\,M_\mathrm{\\odot}$.  The SN
properties were taken from the W7 explosion model \citep{Nomo84} with
an initial orbital separation of $3.75 \times 10^{11}$\, cm.

The temporal evolution of the spatial density distribution of, both,
the companion and the SN ejecta MS\_110 model with and without
including the orbital motion and the rotational velocity of the
companion are shown in Fig.~\ref{Fig:rho} and Fig.~\ref{Fig:rho_rot},
respectively. Some small morphological differences between the rotating and
non-rotating model are seen due to the symmetry-breaking effects
of orbital motion and rotation of the companions star (see Fig.~\ref{Fig:rho_rot}).

\begin{figure}
\centering
\includegraphics[width=0.45\textwidth, angle=360]{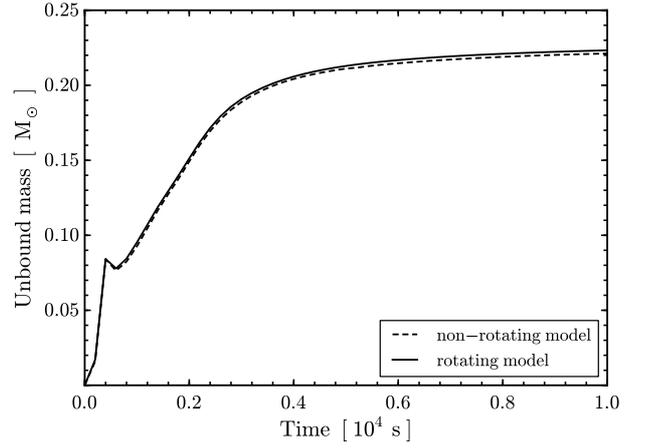}
\caption{Unbound mass from time of explosion with (solid line) and
         without (dotted line)the orbital rotation and spin of 
         of the companion star for model MS\_110. 
         }
\label{Fig:mass}
\end{figure}

\begin{figure*}
\centering
\includegraphics[width=0.45\textwidth, angle=360]{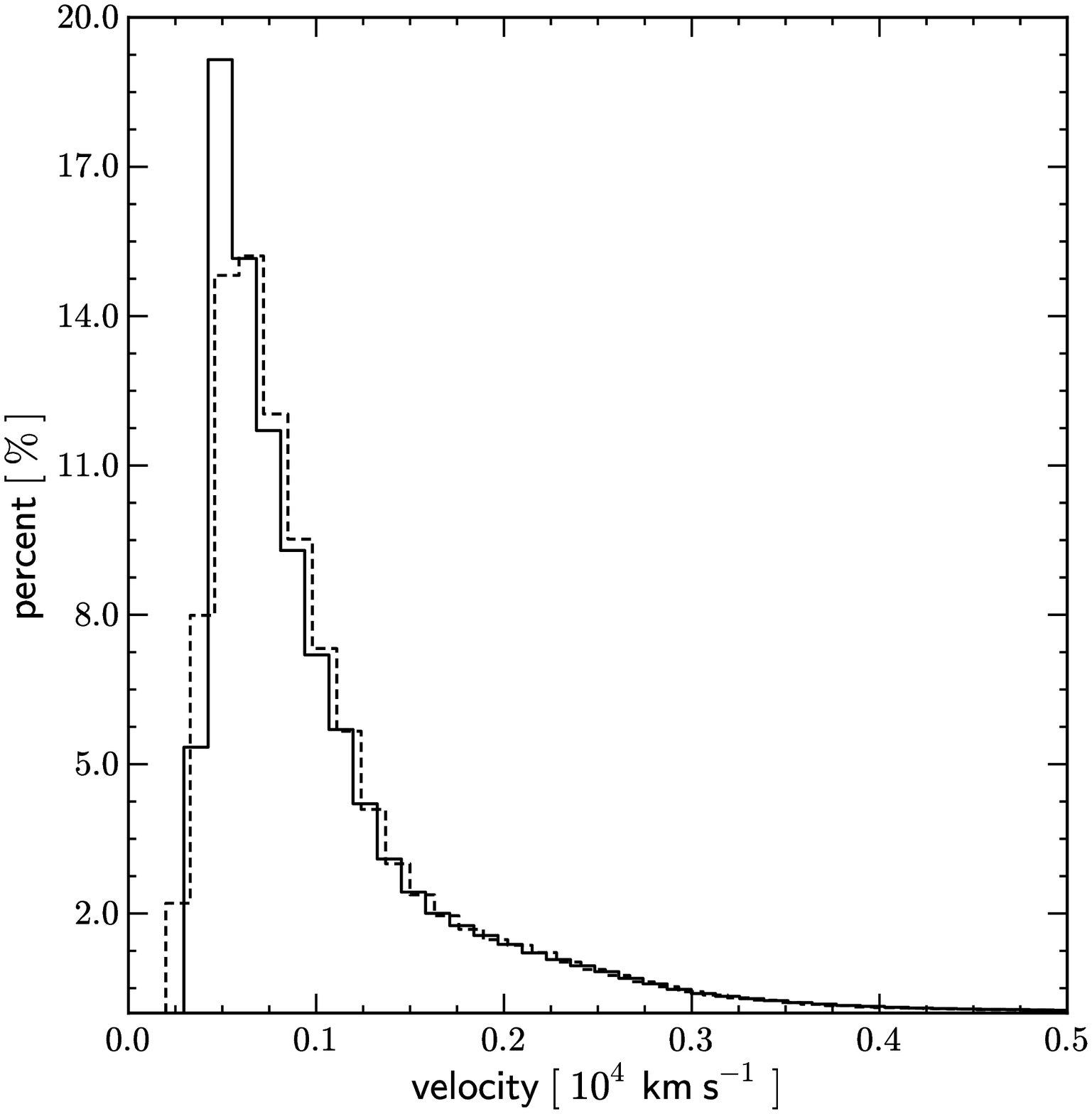}
\includegraphics[width=0.45\textwidth, angle=360]{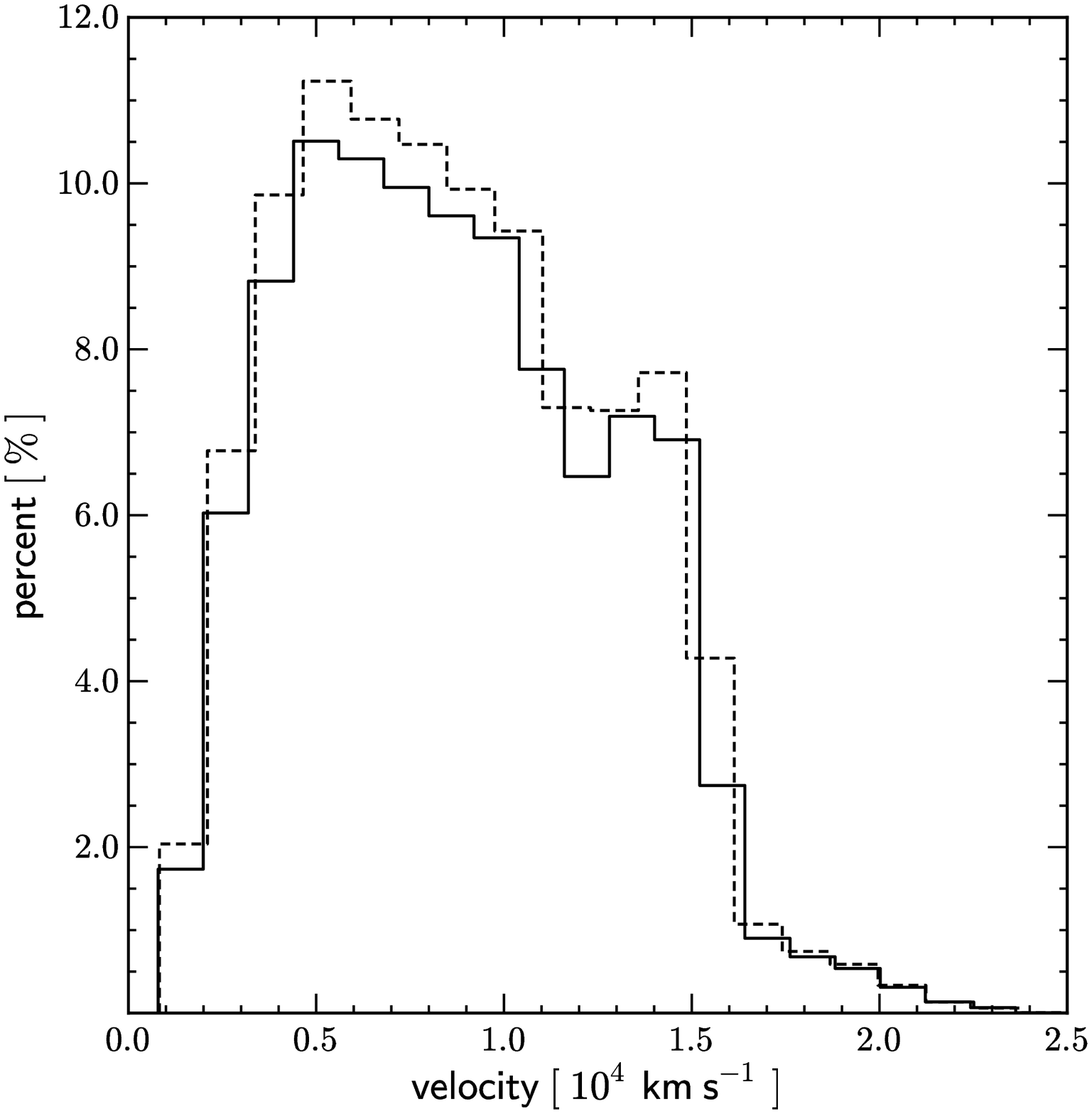}
\caption{Velocity distribution of stripped material that originally belonged to
         the companion star (left figure) and SN ejecta (right figure) for the
         non-rotating (solid lines) and rotating (dashed lines) MS\_110 model. 
         }
\label{Fig:vel_step}
\end{figure*}

At the end of the simulation, a mass of $\sim 0.23\,\mathrm{M}_{\odot}$
($\sim 19\%$ of the total companion mass) is stripped from the
companion star in the MS\_110 model (see Fig.~\ref{Fig:mass}). The
orbital motion and spin of the companion star only leads to a $\sim
2\%$ larger unbound mass (see Fig.~\ref{Fig:mass}), but basically the
same kick velocity ($\sim 60\,\mathrm{km\,s^{-1}}$). The result
differs from what was reported by \citet{Pan12b}, who found that
$16\%$ more mass can be stripped if the orbital motion and spin is
included. In Section~\ref{sec:comparison}, again, we calculate the
amount of unbound mass for the model MS\_160 which has a higher
rotational velocity of $160\,\rm{km\,s^{-1}}$, but still only a $4\%$
difference is found.  Moreover, we do not find that the rotation of the
companion star significantly affects the post-impact velocity
distribution of unbound material and SN Ia ejecta (see
Fig.~\ref{Fig:vel_step}). Most of the stripped hydrogen-rich material
is confined and hidden close to the center of the SN ejecta. The
detection of a hydrogen line may be possible only when the high 
velocity SN ejecta becomes transparent.

To summarize, the main results (e.g., the unbound mass, the kick
velocity) are similar to our previous work (see \citealt{Pakm08,Liu12}) 
even if the orbital and spin velocity of the secondary are taken
into account. This is not surprising, since the orbital and spin
velocity of the companion star are obviously much lower than the
typical values of the expansion velocity of the SN Ia ejecta ($\sim
10^{4}\,\mathrm{km\,s^{-1}}$). Therefore, rotation cannot affect the
basic physics of interactions between the SN ejecta and the companion
star significantly.

\section{Post-impact rotation}
\label{sec:rotation}

In this section, we investigate how the SN Ia explosion affects the
rotation of the companion star in our SPH impact simulations by using
again the MS\_110 model discussed in detail in Section~\ref{sec:model}
as a typical case.

\subsection{Temporal evolution of the overall rotation}
\label{sec:overall}

At the time of the SN explosion, the companion star is expected to rotate rapidly at the
same frequency as the binary system due to the strong tidal interaction during the RLOF phase.
After the SN impact, however, it is uncertain whether it is still possible to use signatures of
rapid rotation to single out candidates for donor stars. This depends 
on the total amount of angular-momentum 
taken by the stripped material and extreme expansion 
of the donor star due to SN ejecta heating.

\begin{figure}
\centering
\includegraphics[width=0.45\textwidth, angle=360]{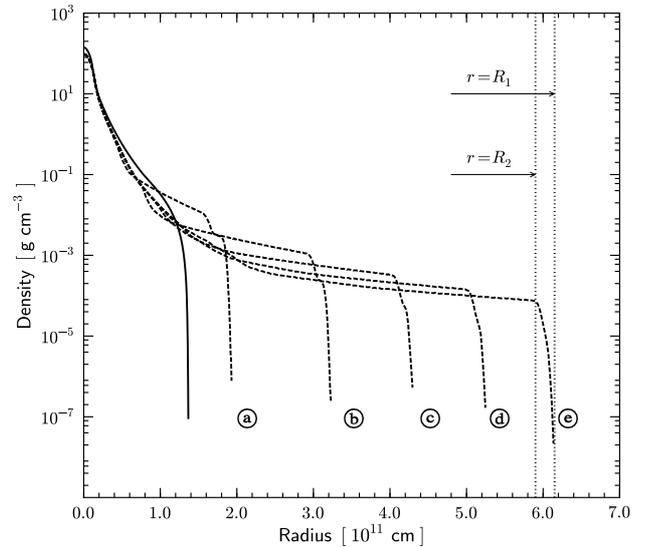}
\caption{Density profile of the MS\_110 model for different times
         after the SN explosion.
         The black solid line corresponds to the initial density 
         profile of the star at the moment of the explosion. All 
         black dashed lines from \textcircled{a} to \textcircled{e} 
         show the radial density distribution of the star 
         $(0.4, 0.8 \cdots 2.0) \times 10^{4}$\,s after the impact 
         in time intervals of $4000\,\mathrm{s}$.}
\label{Fig:raddens}
\end{figure}

By including the orbital motion and spin of the MS\_110 model, we
investigate the total angular momentum that can be carried away by the
stripped mass, studying the post-impact evolution of rotational
velocity of the surviving companion star. During the interaction, the
collision of the SN Ia ejecta brings the companion star out of thermal
equilibrium. Simultaneously, the spherical symmetry of the companion
star is also strongly affected, deforming the shape of the star (the
details of density evolution of a remnant star are shown in
Section~\ref{sec:radial}). Therefore, it is difficult to determine the
real surface of a surviving companion star and to obtain its overall
rotational velocity after the SN impact. For this purpose, the two
following steps are done to estimate the surface of the surviving
companion star in this work.\\
a) We divide the surviving star into
several hundred spherical shells. The density of the SPH particles in
each shell are averaged to calculate a value for that shell. Next, if
the density fluctuation in a shell is too large, it is ignored because
the very outer shells are very poorly resolved anyway (SPH particles
in these shells are too rare to reasonably resolve the structure of
the star). The density profiles of the MS\_110 model as a function of
explosion time are shown in Fig.~\ref{Fig:raddens}. At this stage, we
obtain the surface of surviving star at $R_{1}$, where the tangential
velocity with regard to the initial rotation axis, $z-$axis, is
$v^{\rm{rot}}_{R_{1}}$.\\
b) In Fig.~\ref{Fig:raddens}, a sharp decrease
in density is seen at the outer layer of the surviving companion
star. Here, we simply choose the position $R_{2}$ ($\sim 95\%$ of
$R_{1}$) of the sharp density jump as the real surface of the star
(see the vertical dotted line in Fig.~\ref{Fig:raddens}). Furthermore,
we take the rotation velocity, $v^{\rm{rot}}_{R_{2}}$, at this
surface, $R_{2}$, to denote the overall apparent rotational velocity
of the surviving companion star, $v_{\rm{rot}}$ (i.e., equatorial
rotational velocity). In the following Section~\ref{sec:radial}, it is
shown that the rotational velocity is not changing significantly in the outer
parts of the stellar envelope. The difference between
$v^{\rm{rot}}_{R_{1}}$ and $v^{\rm{rot}}_{R_{2}}$ is $\lesssim
5\%$. On the other hand, it is also difficult to determine the exact
position of the photosphere of the surviving companion star. However,
Figure~\ref{Fig:raddens} shows that the density at the position
$R_{2}$ is always above the lower density-limit of the photosphere of
the sun ($\sim 10^{-7}\, \mathrm{g\,cm^{-3}}$). Therefore, using the
rotational velocity of the star, $v^{\rm{rot}}_{R_{2}}$, to estimate
its observed rotational velocity should be a good approximation.

\begin{figure*}
  \begin{center}
    \subfigure
    {\includegraphics[width=0.45\textwidth, angle=360]{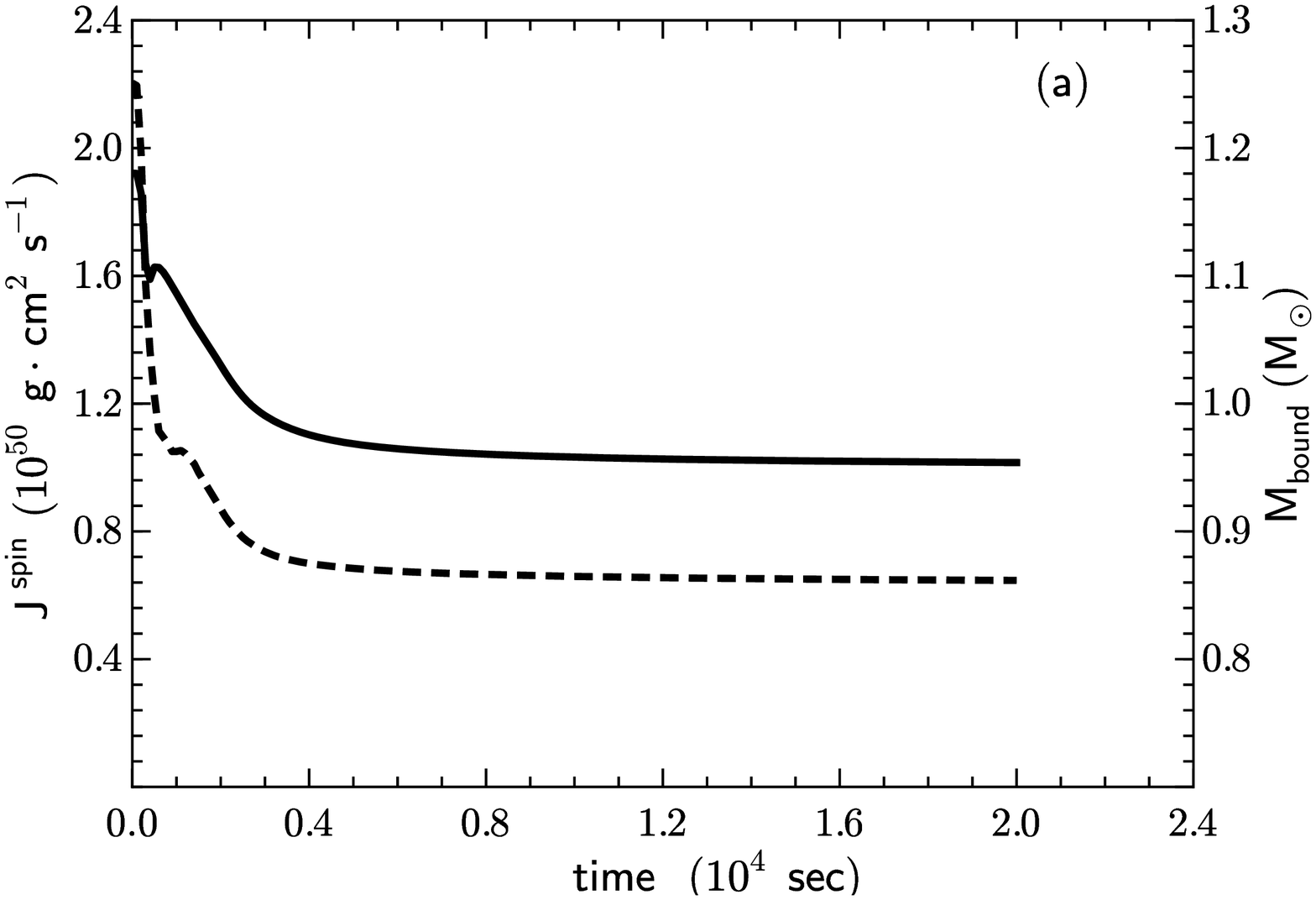}}
    \subfigure
    {\includegraphics[width=0.45\textwidth, angle=360]{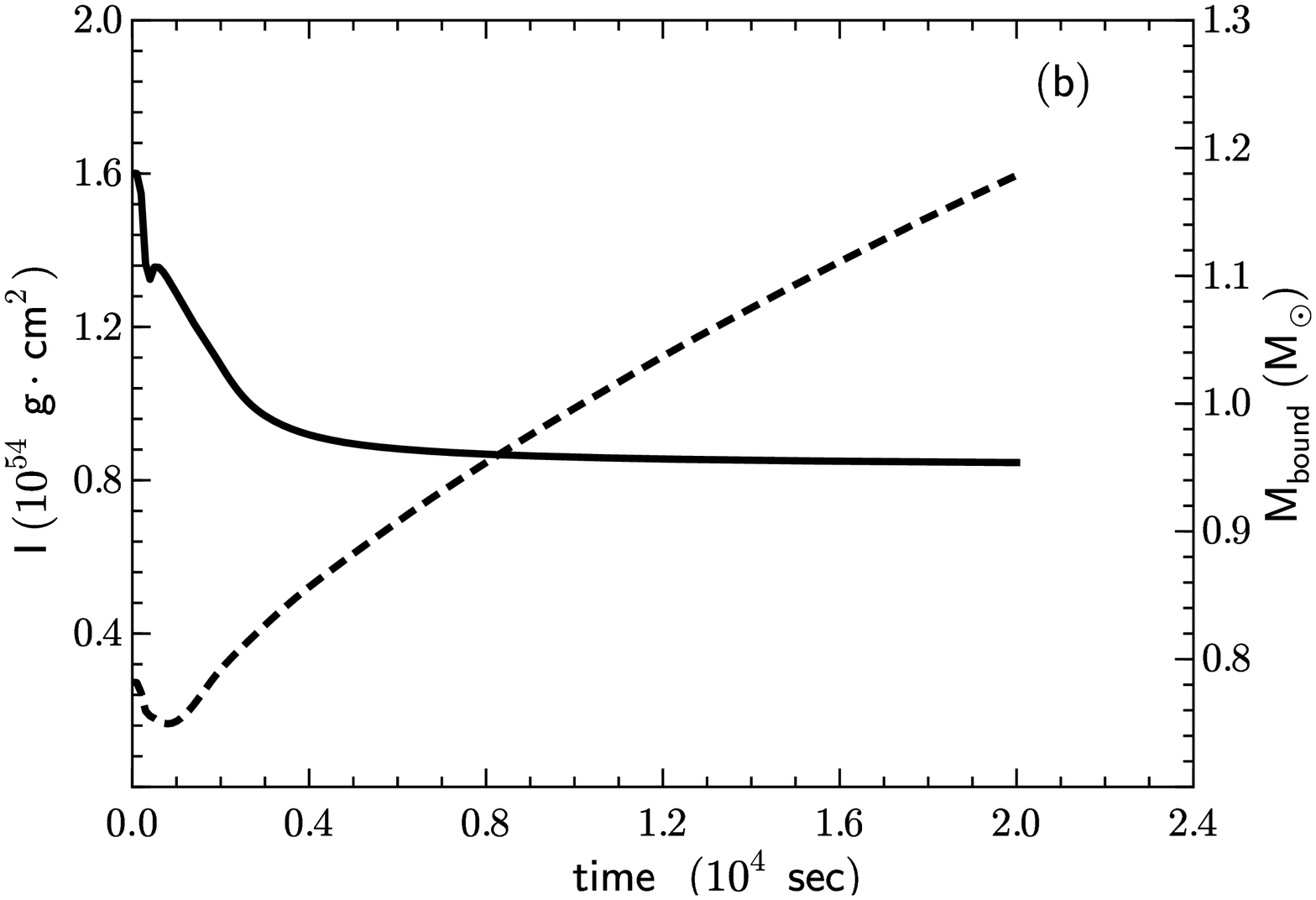}}
    \subfigure
    {\includegraphics[width=0.45\textwidth, angle=360]{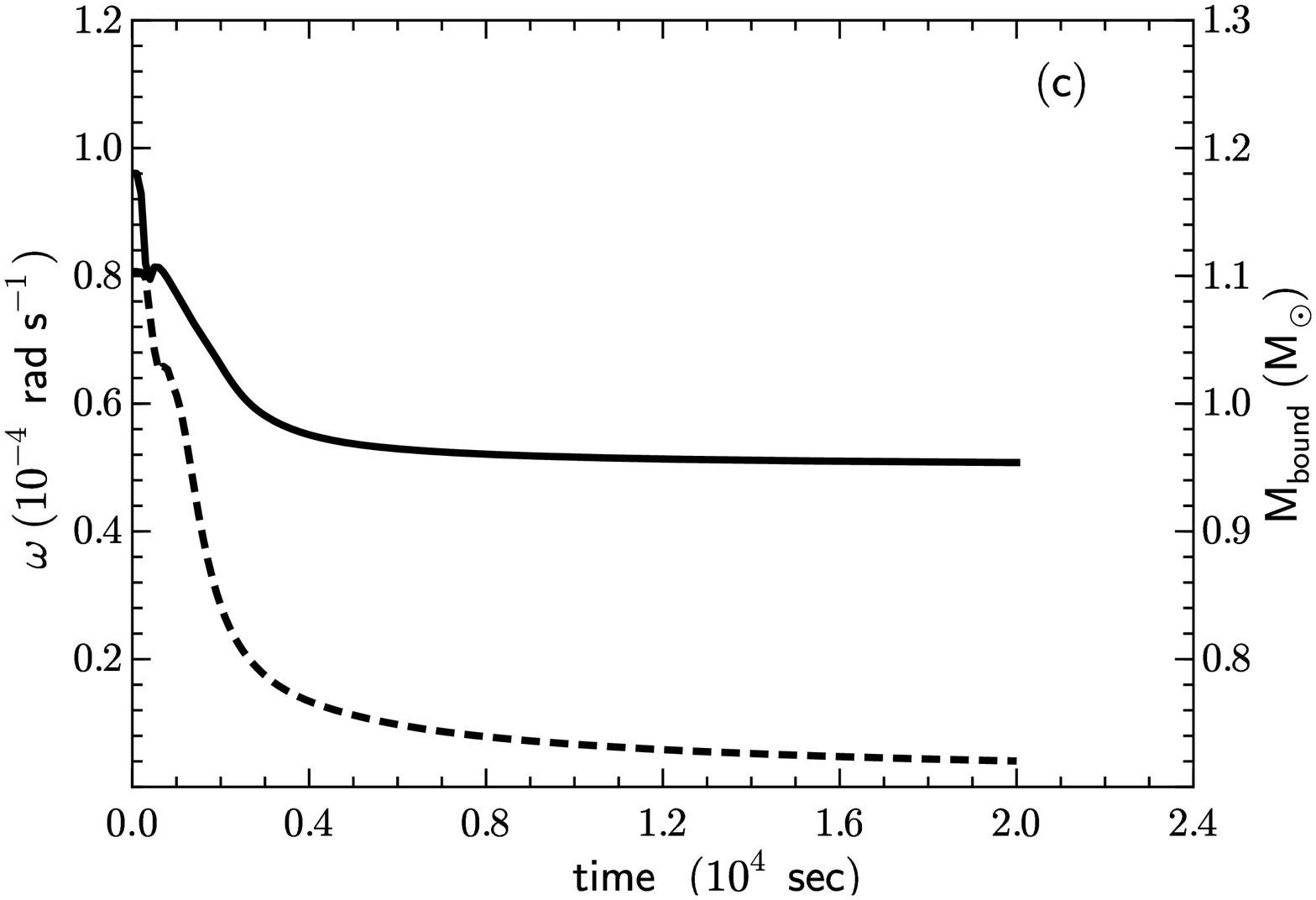}}
    \subfigure
    {\includegraphics[width=0.45\textwidth, angle=360]{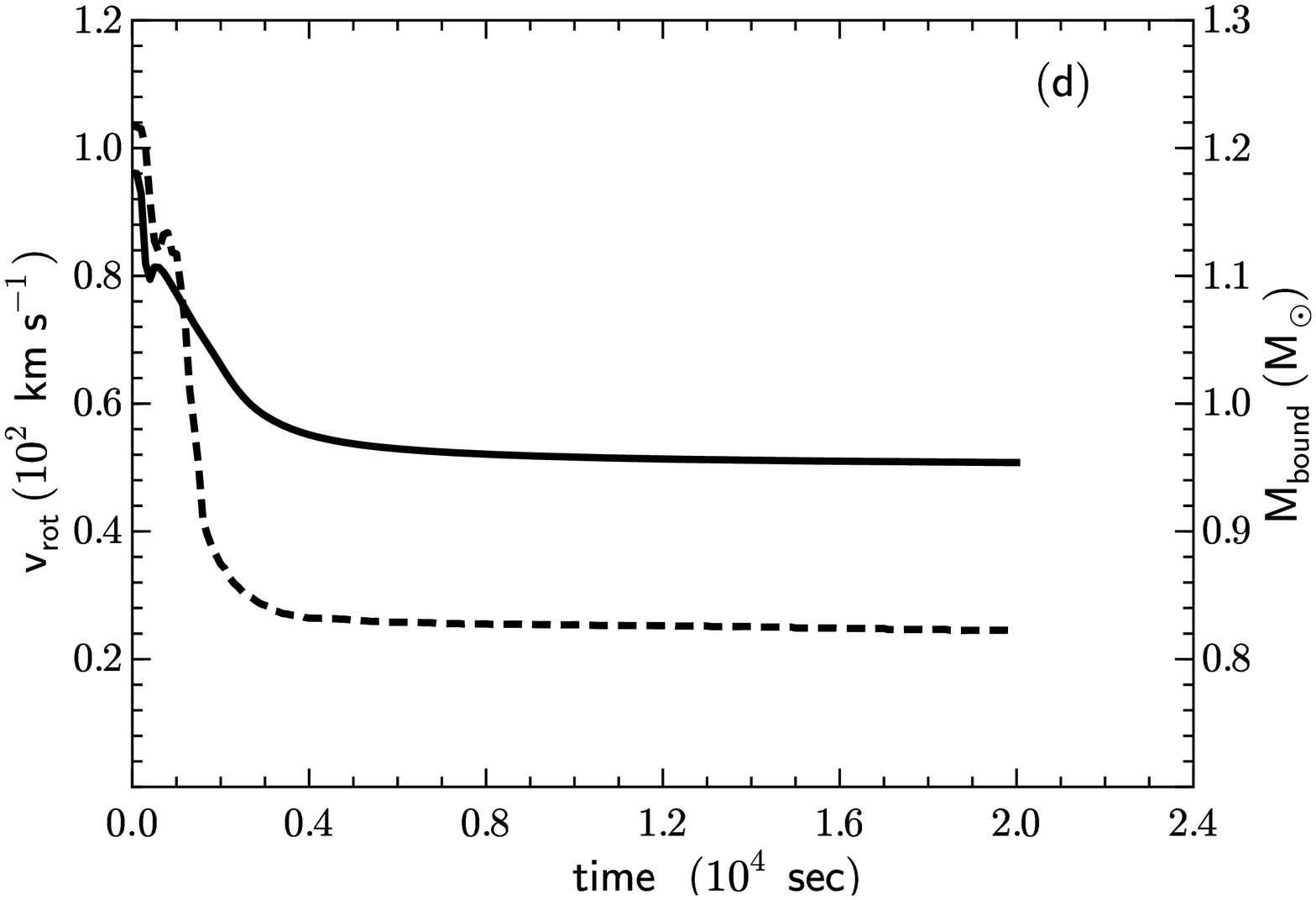}}
\caption{Temporal evolution of some properties of the companion star
         after the SN impact in MS\_110 model. 
         The black solid line corresponds to the total bound mass of
         the companion star, $M_{\rm{bound}}$. The dashed curves show 
         the temporal changes of the total angular momentum
         ($J^{\rm{spin}}$), the moment of inertia ($I$), the surface 
         angular velocity ($\omega$), and the surface rotational
         velocity ($v_{\rm{rot}}$) of the companion star, respectively.}
\label{Fig:overall}
  \end{center}
\end{figure*}

Figure~\ref{Fig:overall} shows the temporal evolution of
$J^{\mathrm{spin}}$, $I$, $\omega$, $v_{\mathrm{rot}}$ and
$M_{\mathrm{bound}}$ of the companion star in the impact simulation
for the MS\_110 model.  Here, $J^{\mathrm{spin}}$, $I$, $\omega$,
$v_{\mathrm{rot}}$ and $M_{\mathrm{bound}}$ are the total angular
momentum, the total moment of inertia, the angular velocity and the
rotational velocity at the equator, and the bound mass of the companion star,
respectively. In the MS\_110 model, the companion star has an initial
total angular momentum of
$2.25\times10^{50}\,\mathrm{g\,cm^{2}\,s^{-1}}$ at the moment of the
explosion. At the end of the simulation, $0.23\,\mathrm{M}_{\odot}$
($\sim 19\%$ of the initial mass) are stripped-off due to the SN
impact, carrying a spin angular momentum of $1.63 \times
10^{50}\,\mathrm{g\,cm^{2}\,s^{-1}}$ ($\sim 72\%$ of that before the
SN explosion). Therefore, the surface rotational velocity of the star is
significantly reduced to $\sim$~25\,$\mathrm{km\,s^{-1}}$ from an
initial rotational velocity of $\sim$~110\,$\mathrm{km\,s^{-1}}$.

The SN impact takes away both mass and angular momentum, which causes
the total angular momentum of the companion star to decrease
sharply. Therefore also the angular velocity and the rotation velocity
(see the dashed lines in Fig.~\ref{Fig:overall}) decrease
significantly.  Furthermore, the total angular momentum stops
decreasing (see dashed line) as the amount of the stripped mass (solid
line) reaches a constant value after $5000~\,\mathrm{s}$, which was
shown in Fig.~\ref{Fig:overall}a. As the companion star is puffed up
due to energy deposition from the impact of the SN ejecta, its outer
layers expand, remarkably increasing the moment of inertia of the star
(see Fig.~\ref{Fig:overall}b). This explains why the angular velocity
keeps slowly decreasing after $5000~\mathrm{s}$ (see dashed line in
Fig.~\ref{Fig:overall}c) although there is no additional mass-loss and
angular-momentum-loss at this moment. However, this expansion does not
reduce the rotational velocity of companion significantly as is
shown by dashed line in Fig.~\ref{Fig:overall}d.

\subsection{Post-impact radial-velocity distribution in the companion star}
\label{sec:radial}

\begin{figure*}
\centering
\includegraphics[width=0.75\textwidth, angle=360]{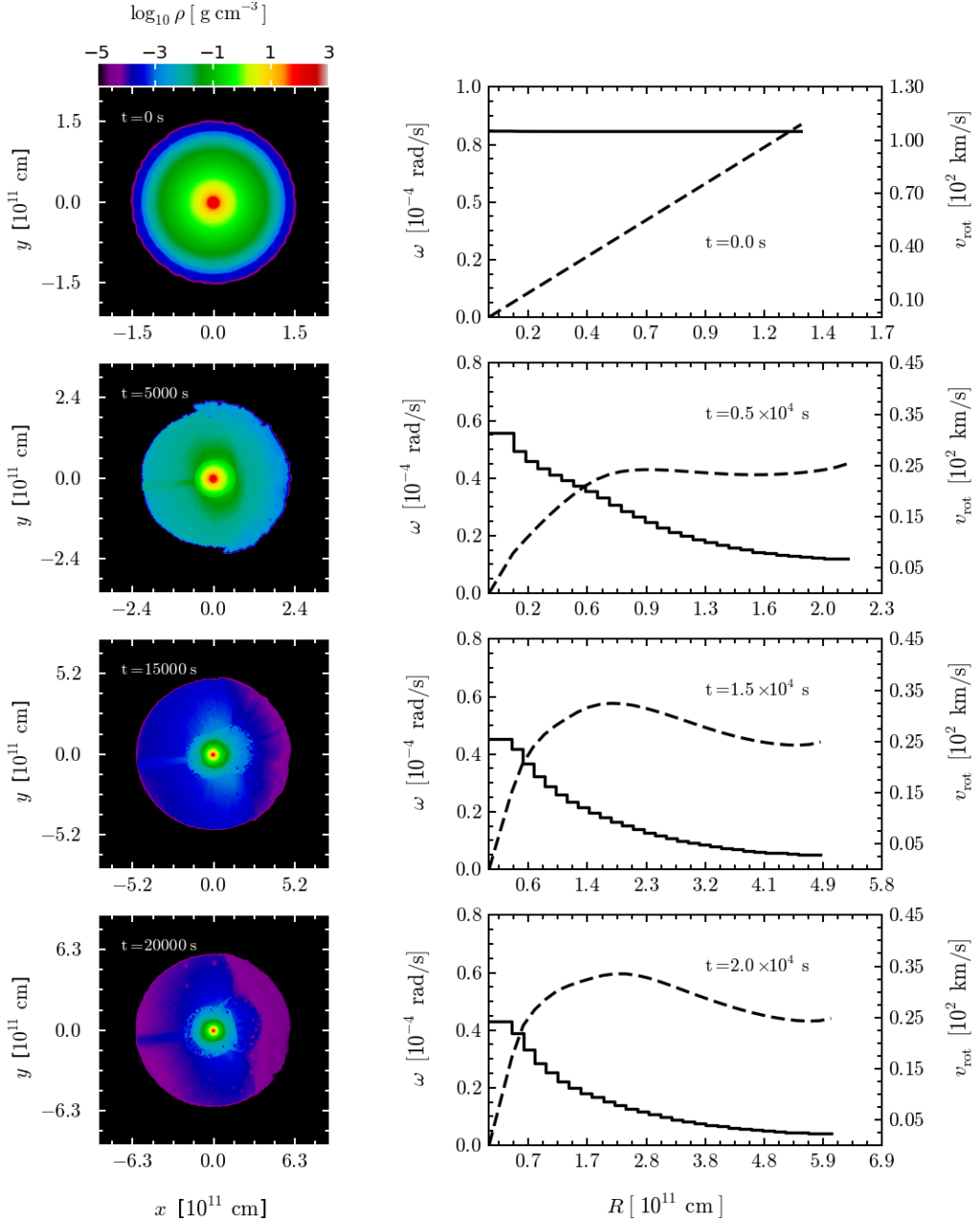}
\caption{ {\it Left column\/}: temporal evolution of the density distribution of the surviving 
          companion star in model MS\_110.  {\it Right column\/}: temporal evolution of the 
          radial configurations of the angular velocity, $\omega$ (solid line), 
          and the rotational velocity, $v_{\mathrm{rot}}$ (dashed
          line). $R$ is the distance from the rotation axes, the
          $z-$axis. Please note that we use different length scales
          in different diagrams.}
\label{Fig:radial}
\end{figure*}

In our simulations, the initial rotation of the companion star was set
up as a rigid-body rotation, the rotation axis being the $z-$axis. At the
moment of the SN explosion, the companion star is spherically
symmetric, having the same angular velocity from the center to the
surface. Therefore, the rotational velocity increases linearly
with radius (see right diagram on the top row of Fig.~\ref{Fig:radial}).

Figure~\ref{Fig:radial} shows the density distribution (left column)
and the radial profiles of the angular velocity (solid curve),
$\omega$, and the rotational velocity (dashed curve),
$v_{\mathrm{rot}}$, of the surviving companion star at different times
after the explosion for the MS\_110 model. After the SN impact,
heating by the SN ejecta puffs up the companion star, causing
its envelope to expand considerably. This produces a object with a
compact inner core and a low-density outer layer (see left column in
Fig.~\ref{Fig:radial}). The star starts to relax towards a spherical
state about $2 \times 10^{4}\,\mathrm{s}$ after the impact. However,
the density in the outer layers varies by as much as a factor of $\sim 3$
in different directions (see bottom row in Fig.~\ref{Fig:radial}),
which was also seen in previous simulations (see \citealt{Mari00, Pakm08, Liu12}). At this moment,
however, the sound crossing time is still longer than the time
we follow in the simulations ($\sim 5.6$\,hours). Therefore, the star
does not reach a spherical state.

\begin{figure}
\centering
\includegraphics[width=0.45\textwidth, angle=360]{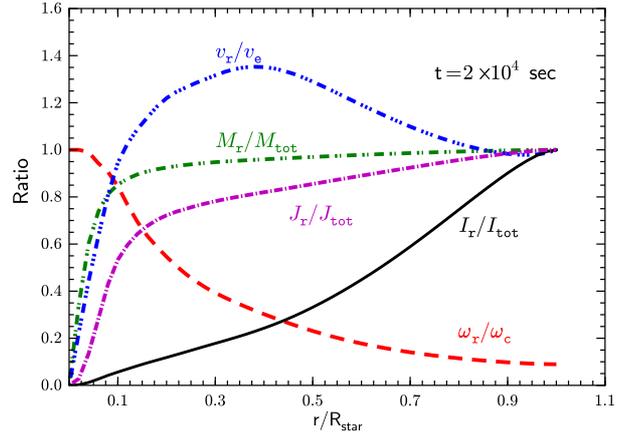}
\caption{Radial rotation profiles of the MS\_110 model $2 \times
         10^{4}$s after the explosion. $R_{\mathrm{star}}$ is 
         the equatorial surface radius and $v_{\mathrm{e}}$ is 
         the equatorial rotational velocity. 
         $M_{\mathrm{r}}$, $J_{\mathrm{r}}$, $I_{\mathrm{r}}$  
         and $\omega_{\mathrm{r}}$ are the total bound 
         mass, the total angular momentum, the angular 
         velocity and the moment of inertia of all bound 
         particles interior to the corresponding cylindrical 
         surface, $r$, respectively. $v_{\mathrm{r}}$ is the 
         circular velocity at distance $r$ from the 
         rotation axis. $\omega_{\mathrm{c}}$ is the
         angular velocity on the rotation axes ($z-$axis).}
\label{Fig:all_r}
\end{figure}

The right column of Fig.~\ref{Fig:radial} shows that the equatorial
rotational velocity of the companion star drops to $\sim
\mathrm{25\,km\,s^{-1}}$ from the original rotational velocity of
$\mathrm{110\,km\,s^{-1}}$ due to the extreme expansion of the star
and the angular momentum losses. The angular velocity near the center
of the star also decreases below its initial value of $0.8
\times 10^{-4}\,\mathrm{rad\,s^{-1}}$ due to the effect of the shock
running through the center, but it becomes stable at late times (see
Fig.~\ref{Fig:radial4}). Additionally, Fig.~\ref{Fig:radial} shows how
the radial rotation profiles of the companion star are affected during
the collision with the SN Ia ejecta (see also
Fig.~\ref{Fig:radial4}). After the SN impact, the companion star is no
longer in rigid-body rotation and its outer layers exhibit some
features of differential rotation. At the equator the companion star
rotates at a different angular velocity than at higher latitudes. The
detailed radial distribution of various rotational properties of model
MS\_110 $2 \times 10^{4}$\,s after the explosion are shown in
Fig.~\ref{Fig:all_r}. The angular velocity decreases with increasing
latitude of the surviving companion star: the equator rotates with a
period of $\sim 14\,\mathrm{days}$, near the poles it is as small as
$2\,\mathrm{days}$. In order ensure that changes of the rotational
profile of the companion star are purely caused by the impact of the
SN explosion, we have also done a test run without the SN. It shows
that without the SN the rotational velocity profile of a companion
star does not change at all over the time of the simulation
($2\times 10^4\,\mathrm{s}$). In our simulation, we use the rotational
velocity at the surface of the post-impact remnant star as the 
potentially observable velocity. If the photosphere of star should be
located at a smaller radius rather than the surface, the rotational
velocity would be a little bit higher than 
$\sim \mathrm{25\ km\,s^{-1}}$. However, the differences of the rotational
velocity are not significant for a large range of outer layers (see
Fig.~\ref{Fig:radial} and Fig.~\ref{Fig:radial4}). Also, in this work, 
the initial rotation of the companion star is set up in rigid rotation,
which might be inconsistent with the realistic rotation of a star. A
more realistic rotation profile of the star is required to replace the
rigid rotation used here in forthcoming investigations, although we do
not expect this to change our main results.

\begin{figure}
\centering
\includegraphics[width=0.45\textwidth, angle=360]{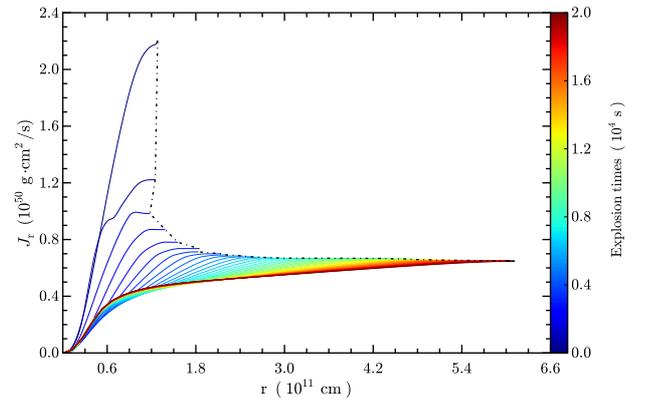}
\caption{Evolution of the distribution of angular momentum inside
         the companion star of model MS\_110. 
         The color scale gives the time since the explosion. 
         Here, $r$ is the distance from the rotation axes ($z$-axis), 
         $J_{\mathrm{r}}$ denotes the angular momentum of all bound 
         particles interior to the corresponding cylindrical surface, 
         $r$. Each colored solid curve shows the radial distribution 
         of angular momentum, $J_{\mathrm{r}}$, at a given time. 
         The dashed-dotted curve corresponds to the temporal evolution 
         of the total angular momentum of the companion star.}
\label{Fig:ang_t}
\end{figure}

Figure~\ref{Fig:ang_t} shows the temporal evolution of the radial distribution 
of the angular momentum of the companion star in model MS\_110. The
different colors belong to 
different times since the SN explosion. Each colored solid 
line shows the details of the radial profile of the angular momentum
at that given time. The angular momentum, $J_{\mathrm{r}}$, was 
calculated by summing up the total angular momentum of all bound 
particles interior to the corresponding cylindrical surface at radius $\mathrm{r}$.
The dashed-dotted line shows the change in total angular momentum as
time progresses. Most of the angular momentum of the surviving 
companion star is concentrated in a small inner core. Outside this 
core, however, only a small fraction of the total angular momentum 
is distributed over an extended outer layer (see  Fig.~\ref{Fig:all_r}).

\section{Comparison with Tycho G}
\label{sec:comparison}

Tycho's SN (SN 1572) is one of only three 
historical SNe Ia observed in our galaxy. SN 1572 has the advantage that 
the field of SNR is not so crowded with stars and, therefore, it provides a
good opportunity to observationally identify the companion in case 
of a SD progenitor. Here, we examine the viability of the 
candidate Tycho G as the possible surviving companion 
star in SN 1572 by comparing its observed rotational velocity 
of $\sim 6 \pm 1.5\,\mathrm{km\ s^{-1}}$ \citep{Kerz09, Kerz12}
with the results of our hydrodynamical simulations.

\begin{table*}
 \centering
 \begin{minipage}{160mm}
  \caption{SPH impact simulations for four different MS companion models.}\label{table:1}
  \begin{tabular}{@{}llrrrrlrlrll@{}}
  \hline\hline
  Model name & $M_{\mathrm{2}}^{\mathrm{SN}}$& $P^{\mathrm{SN}}$&\ \ \ \  $R_{\mathrm{2}}^{\mathrm{SN}}$&\ \ \ \ $a^{\mathrm{SN}}$& $v_{\mathrm{rot}}^{\mathrm{SN}}$\ \ \ & 
  $v_{\mathrm{rot}}^{\mathrm{f}}$\ \ \ & $v_{\mathrm{rot}}^{\mathrm{ff}}$\ \ \ &\ $J_{\mathrm{spin}}^{\mathrm{SN}}$ &\ \ \ \ \ \ \ \ $J_{\mathrm{spin}}^{\mathrm{f}}$&\ \ \ \   $M_{\mathrm{bound}}$ \\
    & $[\mathrm{M}_{\odot}]$ & $\mathrm{[days]}$ &\ \ \ \  $[\mathrm{R}_{\odot}]$ &\ \ \ \  $[\mathrm{R}_{\odot}]$ & \multicolumn{3}{c}{$[\mathrm{km\,s^{-1}}$]}& 
   \multicolumn{2}{c}{[10$\mathrm{^{50}\ g\,cm^{2}\,s^{-1}}$]} &\ \ \ \  $[\mathrm{M}_{\odot}]$\\ 
 \hline
  MS\_160 & 1.21 & 0.29 &\ \ \ \ 0.93 &\ \ \ \ 2.55 & 160\ \ \ & 52\ \ \ & 98\ \ \ & \ \ \ \  2.94  & \ \ \ \ \ \ \ \ 1.31 &\ \ \ \  1.04 \\
  MS\_131 & 1.23 & 0.56 &\ \ \ \ 1.45 &\ \ \ \ 3.94 & 131\ \ \ & 40\ \ \ & 78\ \ \ & \ \ \ \  2.07  & \ \ \ \ \ \ \ \ 0.92 &\ \ \ \  1.06 \\
  MS\_110 & 1.18 & 0.91 &\ \ \ \ 1.97 &\ \ \ \ 5.39 & 110\ \ \ & 25\ \ \ & 46\ \ \ & \ \ \ \  2.25  & \ \ \ \ \ \ \ \ 0.62 &\ \ \ \  0.95 \\
  MS\_081 & 1.09 & 2.00 &\ \ \ \ 3.19 &\ \ \ \ 8.92 & 81\ \ \  & 12\ \ \ & 16\ \ \ & \ \ \ \  2.32  & \ \ \ \ \ \ \ \ 0.26 &\ \ \ \  0.84 \\
\hline
\end{tabular}

\medskip
Here, $M_{\mathrm{2}}^{\mathrm{SN}}$, $P^{\mathrm{SN}}$,
$R_{\mathrm{2}}^{\mathrm{SN}}$, $a^{\mathrm{SN}}$, $v_{\mathrm{rot}}^{\mathrm{SN}}$ 
and $J_{\mathrm{spin}}^{\mathrm{SN}}$ are the mass, the orbital period, the radius,
the spin velocity and angular momentum of the companion star at the
moment of the explosion, respectively.
$v_{\mathrm{rot}}^{\mathrm{f}}$, $J_{\mathrm{spin}}^{\mathrm{f}}$ and
$M_{\mathrm{bound}}$ 
denote the spin velocity, the angular momentum, and
the total bound mass of the companion star after the SN impact. 
$v_{\mathrm{rot}}^{\mathrm{ff}}$ is the rotational velocity 
at the surface after the thermal equilibrium is reestablished. Note
that the rotational velocity, 
$v_{\mathrm{rot}}^{\mathrm{SN}}$,
is calculated by assuming that the rotation of the star is locked 
with the orbital motion due to tidal interactions.  
\end{minipage}
\end{table*}

\begin{figure}
\centering
\includegraphics[width=0.45\textwidth, angle=360]{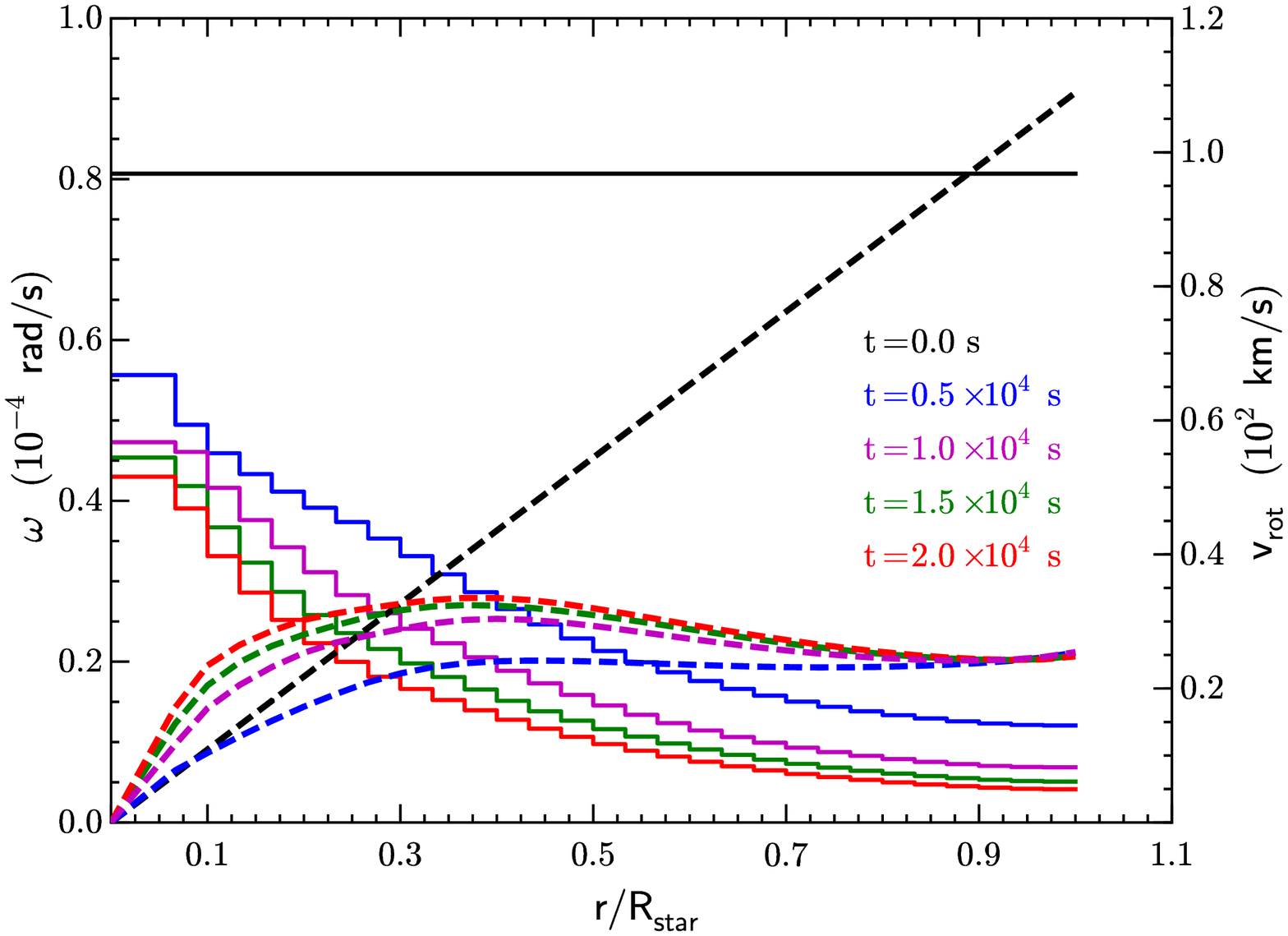}
\caption{Radial profiles of angular velocity (solid curves) and rotational velocity (dashed curves) 
         of the companion star in the MS\_110 model at different times after the impact
         ($\mathrm{0.0, 0.5 \cdots 2.0 \times 10^{4}\,s}$).}
\label{Fig:radial4}
\end{figure}

In our MS\_110 model, after the impact, the rotational velocity of the star
is significantly reduced to $23\%$ of that before the explosion of 
$\sim 110\,\mathrm{km\,s^{-1}}$ (see Table~\ref{table:1}). 
In Section~\ref{sec:overall} we discussed that the surface spin of 
the companion star is roughly constant from $5000~\,\mathrm{s}$ after 
the explosion. However, we stopped our simulation $2 \times 10^{4}$\,s 
after the explosion because of the high computational costs. 
\citet{Pods03} followed the post-impact evolution of the surviving 
companion star further and showed that the star might reestablish 
thermal equilibrium $\sim10^{3}\,\mathrm{yr}$ after the explosion 
(see also \citealt{Pan12a}). Figure~\ref{Fig:radial4} shows that 
the radial distribution of the rotational velocity of the surviving 
companion becomes approximately constant $\sim1 \times 10^{4} \, \mathrm{s}$ after 
the SN explosion, and the surface velocity at the equator is converged 
at the end of our hydrodynamics simulations. Therefore, we can safely assume
that the surface velocity at the equator of the post-impact companion star 
would keep a stable value until its thermal equilibrium is established 
after a few thousand years. Considering that the Tycho SN remnant is only 439
years old, the spin of the MS\_110 model ($\sim 25\,\mathrm{km\,s^{-1}}$) 
after the impact is far larger than the observational rotational
velocity of the star Tycho G ($6 \pm 1.5\,\mathrm{km\,s^{-1}}$  
according to \citealt{Kerz09, Kerz12}).

\citet{Han08} carried out detailed binary evolution calculations for
the WD+MS channel of SNe Ia, in which RLOF starts when the companion
star is on the MS or in the Hertzsprung-gap phase. They obtained many
properties of the companion star at the moment of the SN explosion
(e.g. their masses, spatial velocity, effective temperature,
luminosity, surface gravity, etc). These properties might be verified
by the observations. The distribution of properties of companion stars
in the plane of ($v_{\mathrm{rot}}^{\mathrm{SN}}$,
$M_{\mathrm{2}}^{\mathrm{SN}}$) from \citet{Han08} is shown in
Fig.~\ref{Fig:vrot}a. Based on the observations of \citet{Kerz09,
  Kerz12}, the location of the rotation of the star Tycho G is shown
with black error bars. Clearly, because of the bound rotation before
the impact, Tycho G is far away from the allowed region.

Next we estimate the rotational velocity of the
stars shown in Fig.~\ref{Fig:vrot}a after the impact.  In order to obtain the
distribution of rotational velocities of surviving companions after the
SN Ia explosion, we re-perform 1D consistent binary evolution
calculations to construct the structures of the companions, obtaining
additional three consistent models named ``MS\_081'', ``MS\_131'' and
``MS\_160'' with different rotational velocities of
$81\,\mathrm{km\,s^{-1}}$, $131\,\mathrm{km\,s^{-1}}$ and
$160\,\mathrm{km\,s^{-1}}$ (see Table~\ref{table:1}).  We then use
these three models as input into our impact simulations to investigate
the dependence of the rotational velocity of companion star after the
impact on its value before the explosion. The properties of all
companion models are shown in Table.~\ref{table:1}. We find that to a
good approximation the post-impact rotational velocity of star scales
linearly with the pre-explosion velocity as is shown in
Fig.~\ref{Fig:vel} and can be fitted by 

\begin{equation}
 \label{equation:2}
     V_{\mathrm{rot}}^{\mathrm{f}}= 0.52 \cdot V_{\mathrm{rot}}^{\mathrm{SN}} - 29.8 \ \ \ (\mathrm{km\,s^{-1}}), 
   \end{equation}
where $V_{\mathrm{rot}}^{\mathrm{f}}$ is the post-impact rotational velocity
at the end of the simulations, and $V_{\mathrm{rot}}^{\mathrm{SN}}$ 
corresponds to the star's initial rotational velocity at the moment 
of the explosion.

In the previous work of \citet{Liu12}, it was found that the unbound mass
of the companion star caused by the SN impact is strongly 
dependent on the ratio of separation to the radius of the companion star, 
$a_{\mathrm{f}}/R_{2}$. This relation can be fitted by a power law if  
the effect of the different structures of the companion stars
is neglected (see also \citealt{Mari00, Pakm08, Pan12b}):

   \begin{equation}
    \label{equation:3}
     M_{\mathrm{stripped}}= C_1 \cdot \ \left(\frac{a_{\mathrm{f}}}{R_2}\right)^{\beta} \ \ \mathrm{M}_{\odot},
   \end{equation}
 where $C_1$ is a fitting constant which depends on the different companion star models.
 The parameter $\beta$ is the corresponding power-law index.

Therefore, adopting above power-law relation (see also equation (2)
of \citealt{Liu12}) and the linear relation (\ref{equation:2})
obtained from the data shown in
Fig.~\ref{Fig:vel}, we calculate the final bound mass and the
post-impact rotational velocity of surviving companion stars based on
all companion models from the population synthesis study of
\citet{Han08}. The results are presented in Fig.~\ref{Fig:vrot}b.  The
location of Tycho G is also shown with a black error bars. Here,
we assume that Tycho G is a one solar mass star \citep{Ruiz04} and its
rotational velocity is $6 \pm 1.5\,\mathrm{km\,s^{-1}}$
\citep{Kerz12}.  Fig.~\ref{Fig:vrot} shows that Tycho G  is located
in the outer region of $95.4\%$ of all systems, which casts doubt on
Tycho G star as a promising candidate in SD progenitors of SNe Ia.

However, it is not possible to exclude the star Tycho G completely 
for several reasons.
\begin{enumerate}
  \item The errors shown in Fig.~\ref{Fig:vrot} are based on
    an assumption that Tycho G  is a one solar mass star
    \citep{Ruiz04}. However, it is very difficult to determine the
    actual mass from the observations. Therefore, Tycho G 
    should be located in the vertical gray strip at its given rotation
    velocity in Fig.~\ref{Fig:vrot}b.  If Tycho G is a
    $0.6-0.7\,\mathrm{M}_{\odot}$ star rather a solar mass star, it
    could more likely be a candidate for the companion
    star of SN 1572 in the SD scenario. 
    Moreover, the errors given in
    Fig.\ref{Fig:vrot} are based on an assumption that the
    observed rotational velocity of Tycho G is reduced by an
    inclination angle of $60^{\circ}$. However, if the inclination
    angle is  $30^{\circ}$ instead (see black square in Fig.~\ref{Fig:vrot}b),
    the true rotational velocity of Tycho G would be higher, thus
    making the interpretation that it is the surviving companion
    of SN 1572 more likely. Figure~\ref{Fig:angle} shows the 
    distribution of the observable  $v_{\mathrm{rot}}^{\mathrm{f}}\, \mathrm{sin}\,i$ 
    assuming a random viewing angle.

\begin{figure}
\centering
\includegraphics[width=0.45\textwidth, angle=360]{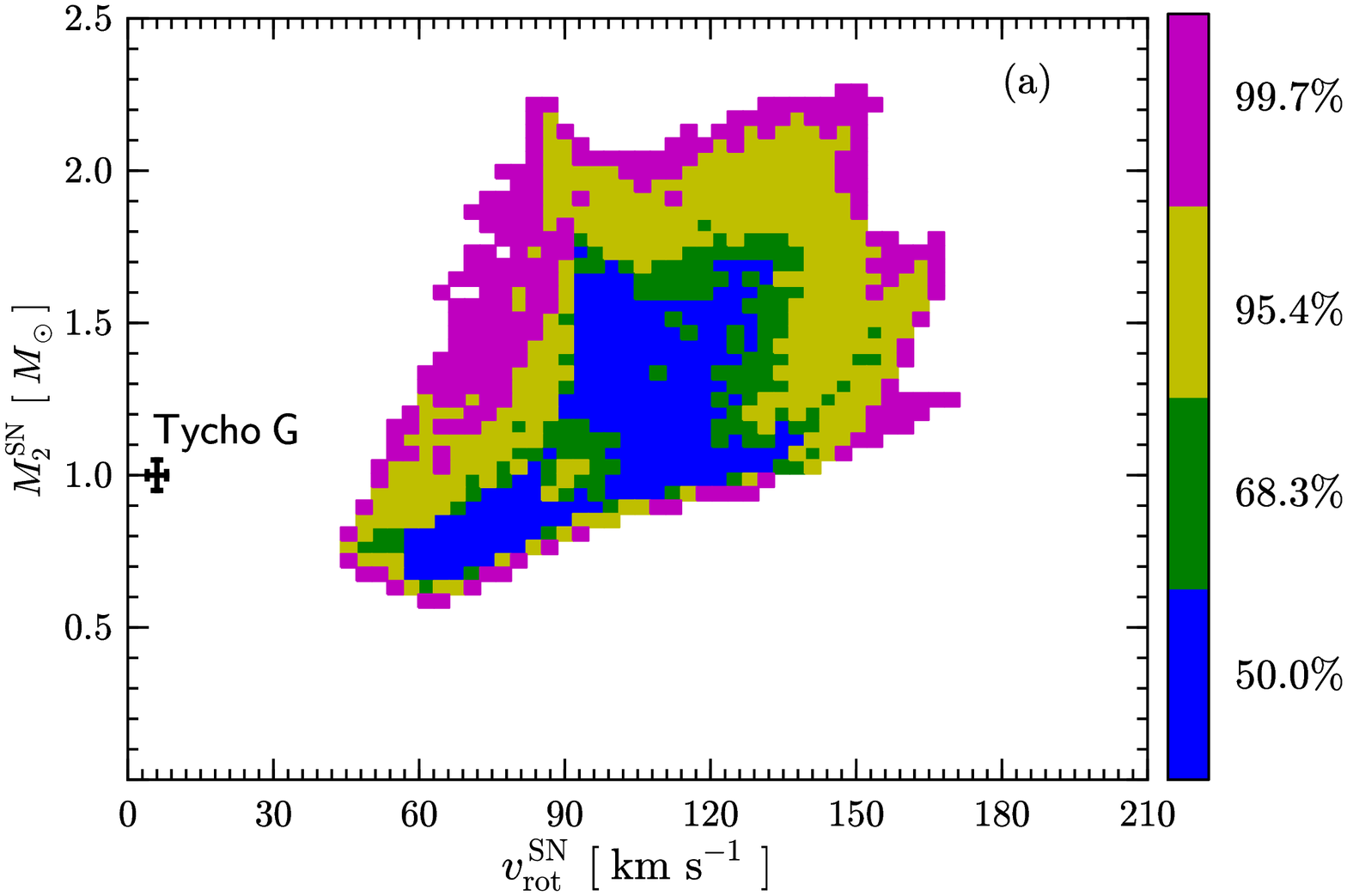}
\includegraphics[width=0.32\textwidth, angle=270]{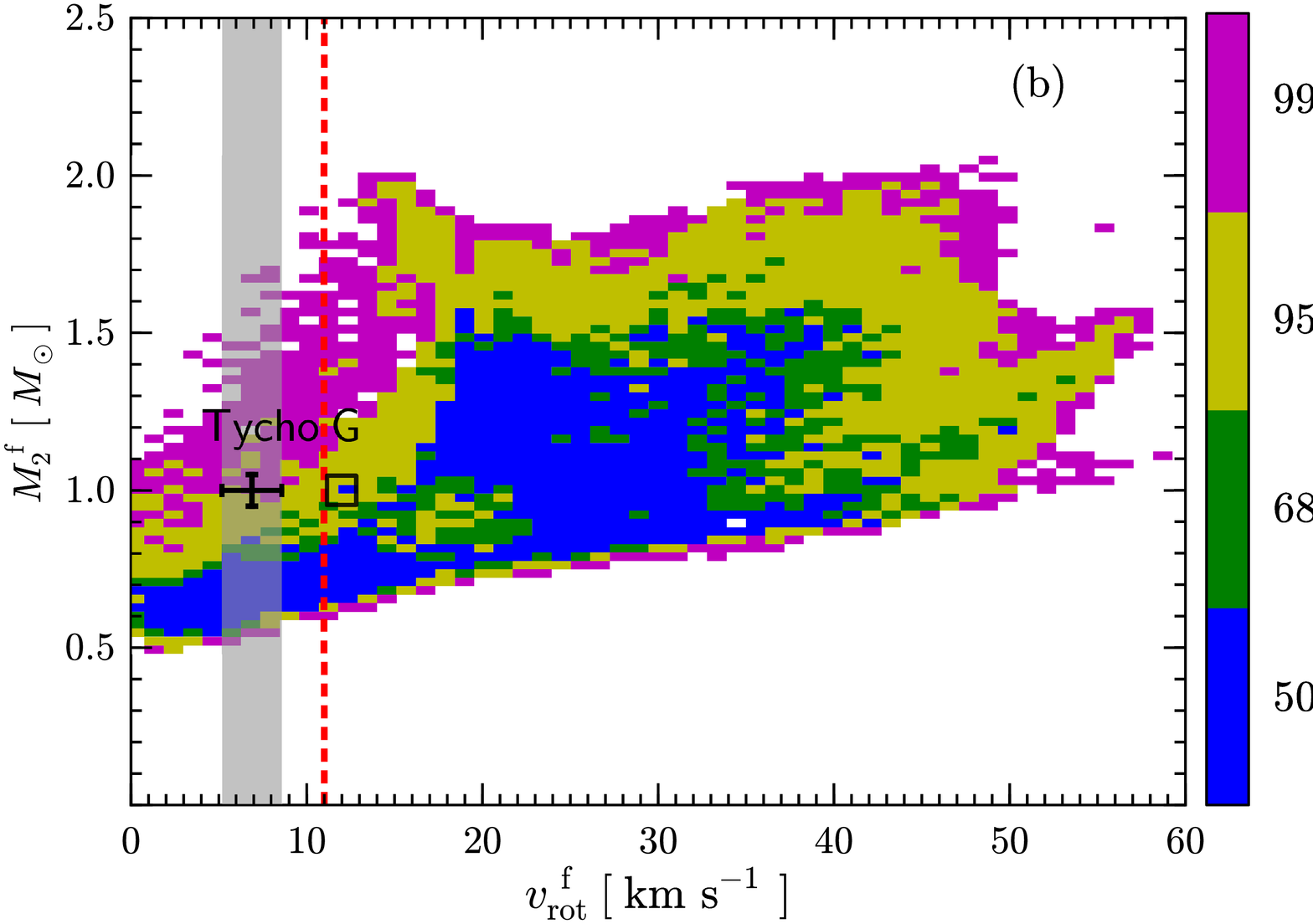}
\caption{{\it Panel (a)\/}: distribution of rotation velocity
         versus mass of the companion stars at the moment of the 
         explosion as obtained by \citet{Han08}. The number of systems 
         decreases from the inner regions to outer ones. The blue area 
         contains $50.0\%$ of all stars, blue plus green $68.3\%$,
         and so on. 
         {\it Panel (b)\/}: as {\it panel (a)\/}, but now the distribution 
         is shown after the impact. The distribution is shifted to
         lower rotational velocity and slightly lower masses. The
         unbound mass stripped by the SN impact is calculated by 
         using equation (2) of \citet{Liu12}, and the post-impact
         rotational velocity of companion star is computed employing 
         the linear relation shown in Fig.~\ref{Fig:vel} (see text). 
         The location of Tycho G according to \citet{Kerz09, Kerz12} 
         is shown with error in black assuming a solar mass 
         star with $5\%$ error \citep{Ruiz04} and an inclination angle 
         of $i=60^{\circ}$. The black square gives the location of 
         Tycho G for an assumed inclination angle of $i=30^{\circ}$.
         The red, vertical dashed line corresponds to the lowest rotational 
         velocity covered by our hydrodynamical simulations (i.e., MS\_081 
         model in Table~\ref{table:1}).} 
\label{Fig:vrot}
\end{figure}

\begin{figure}
\centering
\includegraphics[width=0.45\textwidth, angle=360]{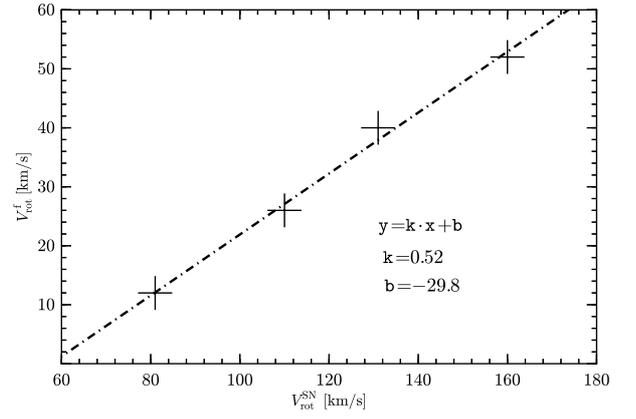}
\caption{Initial rotational velocity at the moment of the SN explosion,
         $V_{\mathrm{rot}}^{\mathrm{SN}}$, versus rotational velocity after the
         impact, $V_{\mathrm{rot}}^{\mathrm{f}}$, for four different MS
         companion star models (see Table~\ref{table:1}).}
\label{Fig:vel}
\end{figure}

  \item In our simulations, we found that the post-impact rotation
    velocity of the surviving companions is linearly dependent on the
    initial rotational velocity at the moment of the explosion (see
    Fig.~\ref{Fig:vel}). However, we restrict our simulations to cover
    a range of initial rotational velocities of only
    $80\,\mathrm{km\,s^{-1}} < v_{\mathrm{rot}}^{\mathrm{SN}} <
    160\,\mathrm{km\,s^{-1}}$ (which corresponds to a range of the
    post-impact rotational velocities of $10\,\mathrm{km\,s^{-1}} <
    v_{\mathrm{rot}}^{\mathrm{f}} < 50\,\mathrm{km\,s^{-1}}$) because
    of computational costs. The companion star models with lower
    initial rotation velocity always evolve to the giant phase with
    larger orbital separation at the time when the supernova explodes,
    which significantly increases the computational effort of the simulations.  Fig.\ref{Fig:vrot}a
    shows that some companion-star models ($\sim 14\%$ of all
    binaries) are located in a range of $v_{\mathrm{rot}}^{\mathrm{SN}} <
    80\,\mathrm{km\,s^{-1}}$ (i.e., regions left of the
    red vertical dashed line in Fig.~\ref{Fig:vrot}b). Actually, these models may not obey the
    linear relation as shown in Fig.~\ref{Fig:vel}, but rather follow
    something close to a power law relation.  Therefore, the models
    with post-impact rotational velocities close to zero in
    Fig.~\ref{Fig:vrot}b might not be realistic. However, $\sim 86\%$
    of all models are located in the range that we follow in the
    simulations ($80\,\mathrm{km\,s^{-1}}$ to $160\,\mathrm{km\,s^{-1}}$),
    indicating that most of post-impact surviving companion stars in
    the WD+MS scenario should be rapidly rotating after the
    impact.

  \item The post-impact masses of the companion stars shown in
    Fig.~\ref{Fig:vrot}b are calculated by directly adopting the power
    law relation from equation (2) of \citet{Liu12}, which ignores
    the effect of the different companion structures. However, \citet{Liu12}
    found that the details of the companion structure also plays an
    important role in determining the amount of unbound mass caused by
    the impact. Nonetheless, the ratio of separation to the
    radius of companion star, $a_{\mathrm{f}}/\mathrm{R_{2}}$, is the
    most important parameter to determine the unbound mass
    \citep{Liu12}.
  \item We assume that the surviving companion star has a
    constant surface rotational velocity after $\sim5000\,\mathrm{s}$
    after the SN Ia impact. However, we only carried out our simulations
    to $2 \times 10^{4}\,\mathrm{s}$ after the explosion. Extended calculations based on the results of our impact
    simulations are still required to follow in detail the
    post-impact rotation of the star during its re-equilibration
    phase. For instance, very recently \citet{Pan12a} showed that the
    remnant star after the SN impact could continue to expand on a
    local thermal timescale of $\sim10^{3}\,\mathrm{yrs}$ before 
    it shrinks again.
\end{enumerate}

\section{Rotation velocity after re-equilibration }
 \label{sec:reestablished}

\begin{figure}
  \begin{center}
    \subfigure
    {\includegraphics[width=0.45\textwidth, angle=360]{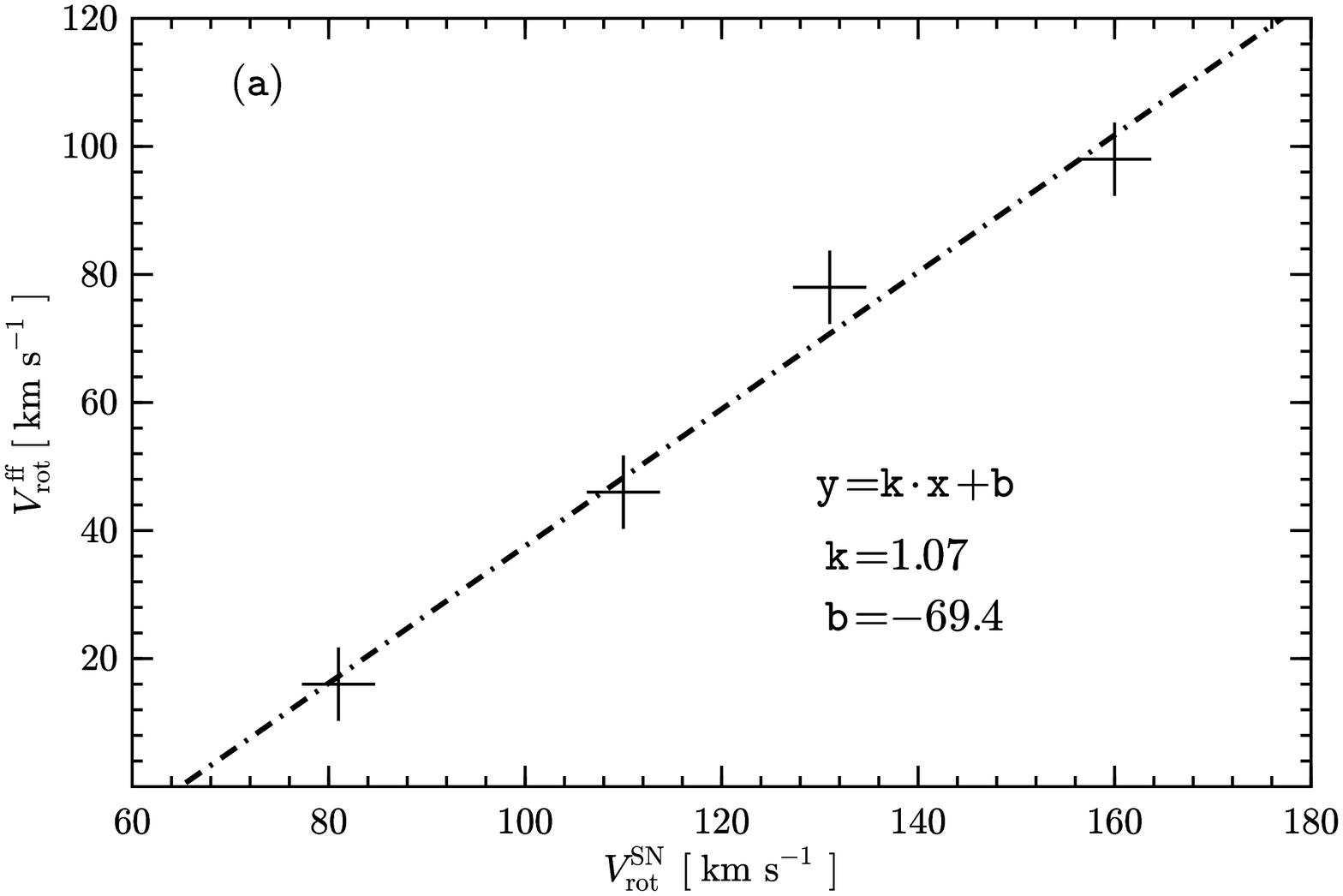}}
    \subfigure
    {\includegraphics[width=0.45\textwidth, angle=360]{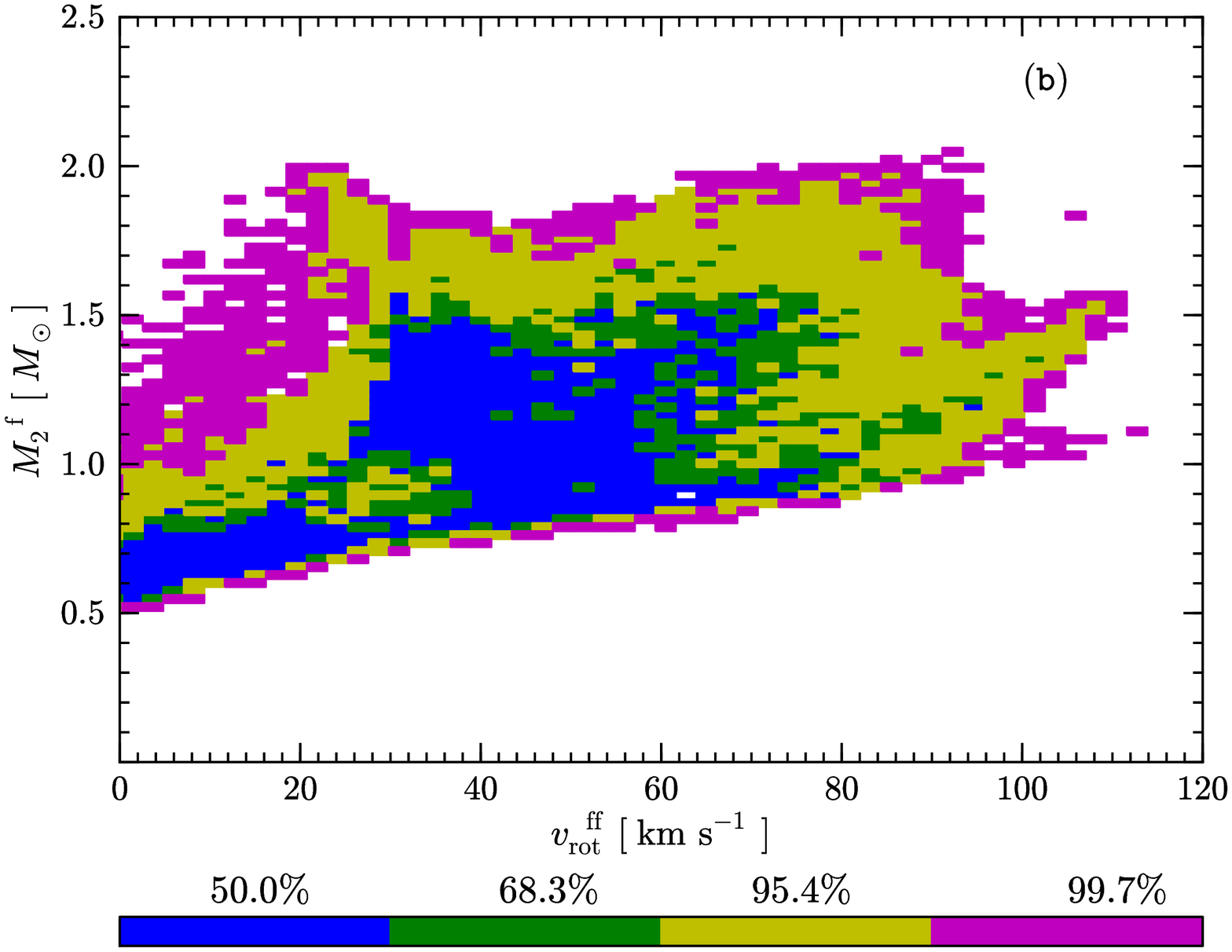}}
\caption{{\it Top panel\/}: similar to Fig.~\ref{Fig:vel}, but in the  plane of 
         ($V_{\mathrm{rot}}^{\mathrm{SN}}$, $V_{\mathrm{rot}}^{\mathrm{ff}}$).
         Note that $V_{\mathrm{rot}}^{\mathrm{ff}}$ is the rotation
         velocity of the remnant star after thermal equilibrium is
         reestablished. {\it Bottom panel\/}: similar to
         Fig.~\ref{Fig:vrot}b, but in the plane of 
         ($V_{\mathrm{rot}}^{\mathrm{ff}}$, $M_{2}^{\mathrm{f}}$). }
\label{Fig:vel1}
  \end{center}
\end{figure}

After the supernova impact, the companion star puffs up and the
stellar envelope is out of thermal equilibrium. The post-impact
remnant star will recover the state of thermal equilibrium on the
Kelvin-Helmholtz timescale of $\sim10^{3}-10^{5}\,\mathrm{yrs}$
\citep{Pan12a, Mari00, Pods03}. In order to estimate the rotation rate
after thermal equilibrium is reestablished, we make the simplifying
assumption that the remnant star has constant rotational velocity during
the re-equilibration phase. After thermal equilibrium is established,
the post-impact remnant star will shrink and return to the state of a
MS-like star. As a result, the surface rotational velocity of the
remnant star will increase again, and we assume that the angular momentum of the star
would redistribute towards rigid-body rotation.  If the angular momentum is
given roughly by $J=\alpha MR^{2}\omega$ before the SN explosion and
after re-equilibration, and also has a constant value of the parameter
$\alpha$, the rotational velocity ($V_{\mathrm{rot}}^{\mathrm{ff}}$)
of the remnant star after the re-equilibration can be determined once
the radius of the re-equilibration star, $R$, is fixed. In order to
estimate the final radius of the remnant star, the MS relation
$R\propto M^{2/3}$ is used (see also \citealt{Mari00}). By adopting
the post-impact angular momentum, $J_{\mathrm{spin}}^{\mathrm{f}}$,
and the value of $\alpha$ that is calculated for the pre-SN companion
star model, we can estimate the rotational velocity of the remnant star
after the thermal equilibrium is reestablished. The rotational velocity of 
different companion-star models before the explosion
and after the re-equilibration can also be fitted with a linear
relation in good approximation (see Fig.~\ref{Fig:vel1}a). 
In Fig.~\ref{Fig:vel1}b we also show the distribution of the 
rotational velocities, $V_{\mathrm{rot}}^{\mathrm{ff}}$, after re-equilibration,
employing the same method as in Section~\ref{sec:comparison}. It can be
seen that the companion stars relax to higher rotation
velocity again after the thermal equilibrium is reestablished (at
least about a few thousand years after the SN explosion).

\begin{figure}
\centering
\includegraphics[width=0.45\textwidth, angle=360]{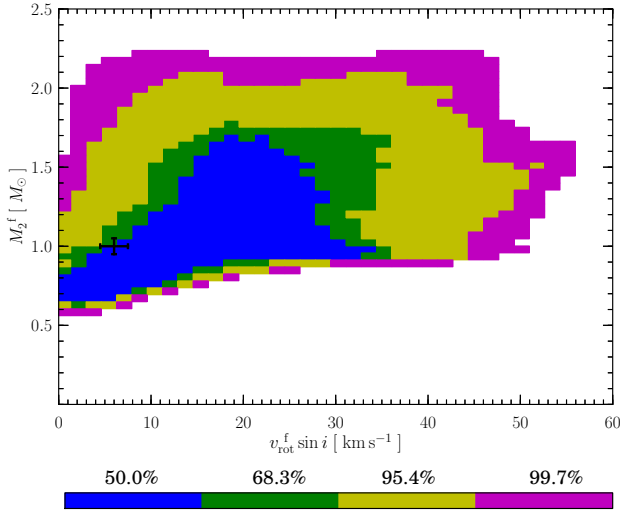}
\caption{Same as Fig.~\ref{Fig:vrot}b, but 
         showing the observable quantity $v_{\mathrm{rot}}^{\mathrm{f}}\, \mathrm{sin}\,i$, 
          where $i$ is the inclination angle.
         The distribution of $i$ is chosen corresponding to a random viewing angle.
         The black cross is 
         again the location of Tycho G with an observed rotation 
         velocity of $6\pm1.5\,\mathrm{km\,s^{-1}}$.}
\label{Fig:angle}
\end{figure}

\section{Summary and conclusions}
\label{sec:conclusions}

In this work, we have modeled the impact of SN Ia ejecta on their MS
companion stars by means of 3D SPH hydrodynamical simulations which
included the orbital and spin velocity of the companion stars. The MS
companion stars were constructed by using Eggleton's 1D stellar
evolution code making use of the optically thick wind model of
\citet{Hach99}. We found that the orbital motion and the spin of the
companion star do not significantly affect the amount of unbound mass
and the kick velocity caused by the SN impact. This result, obtained
with a SPH code, differs from what was found previously by
\citet{Pan12b}, who from their grid-based FLASH code simulations
concluded that these two properties increase the amount of unbound
mass by up to $\sim 16\%$, but also found that the kick velocity is
not affected. We have shown that the SN impact affects the rotation
rate of the companion stars and their rotation laws. In our
simulations we found that the SN impact removes as much as $55\%$ to $89\%$, 
of the initial angular momentum due to the fact
that $14\%$ to $23\%$ of the initial mass is stripped from the MS companion
star. The remnant expands significantly after the impact which causes
the equatorial surface rotational velocity to drop significantly to
$14\%$ to $32\%$ of the original value.  Additionally, we found that the
post-impact rotational velocities of companion stars
depend linearly on those prior to the explosion. 
 Compared with the observed 
 rotational velocity of the presumed  companion star of Tycho's
 supernova, Tycho G,  the final
 rotational velocity we obtain in our simulations is still
 higher by at least a factor of two. Whether or no this difference is
 significant, and may cast doubts on the suggestion that Tycho G is the
 companion of SN 1572, has to be investigated in future studies. In
particular, having a more accurate mass and inclination of Tycho G
would help, as well as simulations that follow the evolution for much
longer than we can do with our explicit SPH code.
Furthermore, by using the
population synthesis results of \citet{Han08}, we showed the
distributions of the rotational velocities of surviving companions
after the impact, which may be very useful for further observations to
identify the surviving companion stars in SNRs.

\section*{Acknowledgments}

      Z.W.L and Z.W.H thank the financial 
      support from the MPG-CAS Joint Doctoral 
      Promotion Program (DP) and Max Planck Institute for Astrophysics (MPA).
      This work is supported by the National Basic Research Program of China 
      (Grant No. 2009CB824800), the National Natural Science Foundation of China (Grant Nos. 11033008 and 11103072) and 
      the Chinese Academy of Sciences (Grant N0. JCS2-WY-T24).
      The work of F.K.R was supported by Deutsche Forschungsgemeinschaft via  the Emmy 
      Noether Program (RO 3676/1-1) and by the ARCHES prize of the German Federal 
      Ministry of Education and Research (BMBF). The simulations were carried out at 
      the Computing Center of the Max Planck Society, Garching,
      Germany.

\bibliographystyle{aa}

\bibliography{ref}


\end{document}